\documentclass[IEEEtran,11pt]{article}
\usepackage{latexsym,mathtools}
\usepackage{graphicx,amsmath,amsfonts,amssymb,color,mathrsfs,placeins,varioref}

\usepackage{epsfig}

 \newcommand{\QED}{\hfill \thicklines \framebox(6.6,6.6)[l]{}}

\numberwithin{equation}{section}

 \newtheorem{definition}{Definition}[section]

\newcommand{\eqnb}{\begin{eqnarray*}}
\newcommand{\eqne}{\end{eqnarray*}}

\def\beqlb{\begin{eqnarray}}\def\eeqlb{\end{eqnarray}}
\def\beqnn{\begin{eqnarray*}}\def\eeqnn{\end{eqnarray*}}

\title{\Large \bf Estimating value at risk and conditional tail expectation for extreme and aggregate risks}
\author{
Suman Thapa   \\
School of Mathematics and Statistics \\
\vspace*{4mm}
Carleton University, Ottawa, ON Canada K1S 5B6 \\
Yiqiang Q. Zhao\\
School of Mathematics and Statistics \\
Carleton University, Ottawa, ON Canada K1S 5B6}
\date{January, 2021}

\begin{document}
\maketitle
\begin{abstract}
In this paper, we investigate risk measures such as value at risk (VaR) and the conditional tail expectation (CTE) of the extreme (maximum and minimum) and the aggregate (total) of two dependent risks. In finance, insurance and the other fields, when people invest their money in two or more dependent or independent markets, it is very important to know the extreme and total risk before the investment. To find these risk measures for dependent cases is quite challenging, which has not been reported in the literature to the best of our knowledge. We use the FGM copula for modelling the dependence as it is relatively simple for computational purposes and has empirical successes. The marginal of the risks are considered as exponential and pareto, separately, for the case of extreme risk and as exponential for the case of the total risk. The effect of the degree of dependency on the VaR and CTE of the extreme and total risks is analyzed. We also make comparisons for the dependent and independent risks.  Moreover, we propose a new risk measure called median of tail (MoT) and investigate MoT for the extreme and aggregate dependent risks.
\vspace*{5mm}\\

\noindent \textbf{Keywords:} Dependent risk measures; median of tail; extreme risks; aggregate risk.
\medskip

\end{abstract}

\section{Introduction}
In this paper, we derive the joint risk measures and extreme and aggregate values of dependent risk measures. In finance and insurance, several risk measures, such as value at risk (VaR), conditional tail expectation (CTE), the distorted risk measure, the copula distorted risk measures, among possible others, are considered, of which VaR and the CTE arae the most popular ones. \\

Let $X$ be a nonnegative random variable with distribution function $F_X(x),$ which represents the risk or claim for an insurance company, or a loss of a portfolio. Let $F_X^{-1}$ be the left continuous inverse of $F_X,$ called the quantile function. For every $\alpha \in [0,1]$, value at risk is denoted by $\text{VaR}_X(\alpha)$ and defined by
$$ \text{VaR}_X(\alpha)= \inf(x: F_X(x) \geq \alpha).$$
In actuarial science, it is also known as the quantile risk measure. VaR is often specified with a confidence level, say $\alpha = 90\%$ or $95\%$ or $99\%$. Hence, $\text{VaR}_X(\alpha)$ represents the loss such that the probability distribution of $\text{VaR}_X(\alpha)$ will not exceed $\alpha$. \\
The conditional tail expectation (CTE), or the expected shortfall of $X$ given that $X > \text{VaR}_X(\alpha)$, denoted by $\text{CTE}_X(\alpha)$ is defined by
$$ \text{CTE}_X(\alpha) = E(X|X > \text{VaR}_X(\alpha)).$$

Both VaR and CTE are important measures for the right tail risk, which are most often studied in insurance and financial investment. CTE satisfies all required properties of a coherent risk measure by \cite{artzner1999coherent}. So, CTE is more preferable than VaR in many applications. However, VaR could be better for optimizing portfolios when good tail models are not available.\\

In practical applications of probability and statistics, the results of an experiment are often  described by more that one random vector that form a multivariate random variable. For example, a person can invest his/her income in more than one market.
Let a multivariate random variable $X = (X_1, X_2,...,X_n)$ be a risk vector, where $X_i$ ($i= 1,2,....,n$) denotes the risk or loss in the sub-portfolio $i$. If people invest their money in different markets, say $X_1,X_2,..,X_n$ and if the markets are not independent, we need to analyze dependent random variables. Copulas are the functions that describe dependencies among variables and provide a way to create distributions to model dependent multivariate data. Using copulas, one can construct a multivariate distribution by specifying marginal distributions.\\

The joint risk measure of $X$ is defined by $ P(X_1 \leq x$, $X_2 \leq x$,...,$X_n \leq x$) and the extreme risks in a portfolio are $X_{(1)}$= $\min(X_1$, $ X_2$,...,$X_n$) and $X_{(n)}$= $\max(X_1$, $X_2,...,X_n)$ consisting of the $n$ sub-portfolios. The extreme risks and the aggregate risk are very important and popular. We study the cases when the risks in the portfolio follow the exponential and pareto distributions. Specifically, we consider two subportfolios, say $X_1$ and $X_2$ of the portfolio, say $X$. Since FGM copula is very popular and easier to use, we apply FGM copula \cite{fischer2007constructing} to analyze the dependency between the two risks.

We derive VaR and CTE for extreme and aggregate risks of two dependent risks. That is, for $X=(X_1, X_2)$, we calculate VaR and CTE for the extreme risks ($X_{(1)}=\min(X_1, X_2)$ and $X_{(2)}= \max(X_1, X_2)$) and the aggregate risk ($X=X_1 + X_2$) when $X_1$ and $X_2$ are dependent.\\

There are some papers in the literature in which copula was used in financial risks. Brahim, Fatah and Djabrane, in their paper \cite{brahim2018copula}, derived a new risk measure called the  copula conditional tail expectation that measures the conditional expectation given that the two dependent losses exceed their value at risks. Heilpern, in his paper \cite{heilpern2011aggregate}, used copula to investigate the sum of dependent random variables. The risk measures, value at risk and the expected shortfall of such sums were calculated in his paper. Cai and Li, in their paper \cite{cai2005conditional}, defined the conditional tail expectation of the aggregated risk and the extreme risk for multivariate phase type distributions, but they did not use copula theory. In other papers related to this field, Hardy introduced risk measures in actuarial applications in paper \cite{hardy2006introduction} and Brazauskas, Jones, Puri and Zitikis estimated the  conditional tail expectation with actuarial applications in their paper \cite{brazauskas2008estimating}. Based on the literature review in this field, it is interesting to estimate value at risk and the conditional tail expectation of the extreme risk and the aggregate risk of two dependent risks explicitly by using copula. This is a completely new method to define dependent risk measures that follow the exponential and pareto distributions.\\

We propose a new risk measure called median of tail (MoT). In some cases, VaR is the worst risk measure in the given confidence interval and it is not coherent. The conditional tail expectation (CTE) is useful and coherent. Sometimes, in the absence of good tail models, CTE cannot be a useful measure. Sarykalin, Serraino and Uryasev \cite{sarykalin2008value} tried to explain strong and weak features of these risk measures and illustrate of them with several examples. Yamai and Yoshiba  \cite{yamai2005value} compared and analyzed value at risk and the conditional tail expectation under market stress. When VaR and CTE are not suitable, median of tail (MoT) can be used.\\

The median is less affected by outliers and skewed data than the mean and is usually a preferred measure of central tendency when the distribution is not symmetrical. Mean and median are approximately close in many distributions. In a skewed distribution, the outliers in the tail pull the mean away from the center towards the longer tail. In that case, the median better represents the central tendency for the distribution. In this paper, we derive MoT for the extreme and the aggregate of two dependent risks in the portfolio and generalize the result in section 4.\\

The rest of the paper is organized as follows. In section 2, we estimate VaR and CTE for the extreme and the aggregate of dependent risks which include VaR and CTE for the exponential distribution in section 2.1, VaR and CTE for extreme risks of the exponential distribution in section 2.2, VaR and CTE for the pareto distribution in section 2.3, VaR and CTE for extreme risks of the pareto distribution in section 2.4 and VaR and CTE for the aggregate risk of the exponential distribution in section 2.5. In section 3, we estimate median of tail (MoT) for the  extreme and the aggregate of dependent risks which include MoT for the exponential and pareto distributions in section 3.1, the MoT for extreme risks of the exponential distribution in section 3.2, MoT for extreme risks of the pareto distribution in section 3.3, MoT for the aggregate risk of the exponential distribution in section 3.4 and final conclusions are made in section 4.

\section{VaR and CTE for extreme and aggregate of dependent risks}
In this section, we estimate value at risk (VaR) and the conditional tail expectation (CTE) for the extreme and the aggregate of two dependent risks. The distribution of the risk measures is considered as exponential and pareto.\\
Let $X$ be an exponential random variable with the cumulative distribution function
$$ F_X(x) = P(X \leq x) = 1 - e^{-\lambda x},~{}~{} x \geq 0. $$
Then, $\text{VaR}_X(\alpha)$ is defined as  $\text{VaR}_X(\alpha)= \inf(x : F_X(x) \geq \alpha),$ where $ \alpha \in [0,1]$. Here, $F_X(x)=  1 - e^{-\lambda x} = \alpha $ gives $ x = - \dfrac{1}{\lambda}   \ln(1 - \alpha)$.\\
So, $ \text{VaR}_X(\alpha)= - \dfrac{1}{\lambda} \ln(1 - \alpha)$ and the CTE is given by
$$ \text{CTE}_X(\alpha) = E(X|X > \text{VaR}_X(\alpha))= \dfrac{1}{1 - F_X(\text{VaR}_X(\alpha))} \int_{\text{VaR}_X(\alpha)}^{\infty} x dF_X(x),$$
where $F_X$ is the distribution function of $X$. Since $X$ is continuous, $F_X(\text{VaR}_X(\alpha))= \alpha. $ \\
Therefore,
$$ \text{CTE}_X(\alpha)= \frac{1}{1 - \alpha} \int_{\text{VaR}_X(\alpha)}^{\infty} x dF_X(x) = \frac{1}{1 - \alpha} \int_{\text{VaR}_X(\alpha)}^{\infty} x \lambda e^{-\lambda x}dx. $$
If \( \text{VaR}_X(\alpha)= Q_{\alpha}\), then
\begin{align*}
\text{CTE}_X(\alpha)&= \frac{\lambda Q_{\alpha}e^{-\lambda Q_{\alpha}}+ e^{-\lambda Q_{\alpha}}}{\lambda(1-\alpha)},
\end{align*}
where $Q_{\alpha}= \dfrac{-\ln(1 - \alpha)}{\lambda}$.\\
Therefore, $ \text{CTE}_X(\alpha)= \dfrac{1}{\lambda} + VaR_X(\alpha).$\\

Let $X$ be a Pareto random variable with the distribution function
$$ F_X(x) = P(X \leq x) = 1 - \left(\frac{x_o}{x}\right)^{\gamma},~{} x_o \geq 0,~{} \gamma >0, ~{}x \geq x_o. $$
Then, $\text{VaR}_X(\alpha)$ is defined as $\text{VaR}_X(\alpha)= \inf(x : F_X(x) \geq \alpha),$ where $ \alpha \in [0,1].$ \\
We have that,
$ F_X(x)= 1 - \Big(\dfrac{x_o}{x}\Big)^{\gamma}= \alpha $ gives $ x = x_o(1 - \alpha)^{-\frac{1}{\gamma}}. $ \\
So, $ \text{VaR}_X(\alpha)= Q_{\alpha}= x_o(1-\alpha)^{-\frac{1}{\gamma}}$ and CTE is given by
$$ \text{CTE}_X(\alpha) = E(X|X > \text{VaR}_X(\alpha))= \frac{1}{1 - \alpha} \int_{\text{VaR}_X(\alpha)}^{\infty} x dF_X(x),$$
where $F_X$ is the distribution function of $X$. Since the probability density function of $X$ is $f_X(x)= \gamma x_o^{\gamma}x^{-\gamma-1},$
$$ \text{CTE}_X(\alpha)= \frac{1}{1 - \alpha}\int_{Q_{\alpha}}^{\infty} x .\gamma. x_o^{\gamma}. x^{-\gamma-1}dx = \frac{\gamma x_o^{\gamma}}{1- \alpha}\left[\frac{x^{-\gamma}}{-\gamma}\right]_{Q_{\alpha}}^{\infty} = \frac{\gamma x_o^{\gamma} Q_{\alpha}^{1-\gamma}}{(\gamma-1)(1-\alpha)}.$$

 Using \( Q_{\alpha}= x_o(1 - \alpha)^{-\frac{1}{\gamma}},\)
 $$ \text{CTE}_X(\alpha)= \frac{\gamma x_o}{(1 - \alpha)^{\frac{1}{\gamma}}(\gamma - 1)}= \frac{\gamma}{\gamma-1} \text{VaR}_X(\alpha).$$

\subsection{VaR and CTE for extreme risks of exponential distribution}
Let $X = (X_1, X_2)$ be a risk vector where $X_1$ and $X_2$ denote the risks in the subportfolio of $X$. Then, $X_{(1)} = \min(X_1, X_2)$ and $X_{(2)}= \max(X_1, X_2)$ are the extreme risks in a portfolio.\\
Let $X_1$ and $X_2$ be $\exp(\lambda_1)$ and $\exp(\lambda_2)$ distributions. Then, the distribution function of $X_i$ ($i=1,2$) is given by $ P(X_i \leq x) = F_{X_i}(x)= 1 - e^{-\lambda_i x}, i= 1, 2. $ \\
The distribution function of $X_{(1)}= \min(X_1, X_2)$ is given by
$$ P(X_{(1)} \leq x) = 1 - P\big(\min(X_1, X_2) > x \big) = 1 - P(X_1 > x , X_2 > x). $$

\textbf{Case (i)} When $X_1$ and $X_2$ are independent, we have
$$ P(X_{(1)} \leq x) = 1 - P(X_1 > x , X_2 > x)= 1 - e^{-\lambda_1 x}\cdot e^{-\lambda_2 x} = 1 - e^{-(\lambda_1+\lambda_2)x}.$$
So, $ X_{(1)} \sim exp(\lambda_1 + \lambda_2). $ Therefore,
$$ \text{VaR}_{X_{(1)}}(\alpha)= - \frac{- \ln(1 - \alpha)}{\lambda_1 + \lambda_2} ~{}~{}
, ~{}~{} \text{CTE}_{X_{(1)}}(\alpha)= \frac{1}{\lambda_1 + \lambda_2} + VaR_{X_{(1)}}(\alpha)= \frac{1 - \ln(1 - \alpha)}{\lambda_1 + \lambda_2}. $$
The distribution function of $ X_{(2)}= \max(X_1, X_2)$ is given by
\begin{align*}
P(X_{(2)} \leq x) &= P\big(\max(X_1, X_2) \leq x\big)= P(X_1 \leq x, X_2 \leq x)= P(X_1 \leq x)P(X_2 \leq x)\\
&= (1 - e^{-\lambda_1 x})(1 - e^{-\lambda_2 x})= 1 - e^{-\lambda_1 x}- e^{-\lambda_2 x}+ e^{-(\lambda_1+ \lambda)x},
\end{align*}
and its density function is given by
$$ f_{X_{(2)}} = \lambda_1 e^{-\lambda_1 x} +  \lambda_2 e^{-\lambda_2 x} - (\lambda_1+ \lambda_2)  e^{-(\lambda_1+ \lambda_2)x}.$$
If $ \text{VaR}_{X_{(2)}}(\alpha) = Q_{\alpha}$, then
$$ F_{X_{(2)}}(x) = 1 - e^{-\lambda_1 Q_{\alpha}}- e^{-\lambda_2 Q_{\alpha}}+ e^{-(\lambda_1+ \lambda)Q_{\alpha}}= \alpha ~{}~{} and $$
$$ \text{CTE}_{X_{(2)}}(\alpha)= \frac{1}{1 - \alpha} \int_{Q_{\alpha}}^{\infty} x dF_X(x)= \frac{1}{1 - \alpha} \int_{Q_{\alpha}}^{\infty} x\Big[\lambda_1 e^{-\lambda_1 x} + \lambda_2 e^{-\lambda_2 x} - (\lambda_1+ \lambda_2)  e^{-(\lambda_1+ \lambda_2)x}\Big]dx. $$
After calculations, we have
\begin{equation}
\text{CTE}_{X_{(2)}}(\alpha) =\frac{\lambda_1 Q_{\alpha} e^{-\lambda_1 Q_{\alpha}} + e^{-\lambda_1 Q_{\alpha}}}{\lambda_1(1- \alpha)} +  \frac{\lambda_2 Q_{\alpha} e^{-\lambda_2 Q_{\alpha}}+ e^{-\lambda_2 Q_{\alpha}}}{\lambda_2(1- \alpha)} \nonumber - \frac{(\lambda_1+\lambda_2) Q_{\alpha} e^{-(\lambda_1+\lambda_2)Q_{\alpha}}+ e^{-(\lambda_1+\lambda_2)Q_{\alpha}}}{(\lambda_1+\lambda_2)(1- \alpha)}.
\end{equation}

\textbf{Examples:}
If we choose $\lambda_1 = 0.5$, $\lambda_2 = 0.6 $ and $\alpha = 0.9$, then
\begin{align*}
\text{VaR}_{X_1}(0.9) &= \frac{-\ln(1 - 0.9)}{0.5} = 4.605,~{} \text{VaR}_{X_2}(0.9) = \frac{-\ln(1 - 0.9)}{0.6} = 3.837, \\
\text{CTE}_{X_1}(0.9) &= \frac{1 - \ln(1 - 0.9)}{0.5} = 6.605, ~{} \text{CTE}_{X_2}(0.9) = \frac{1 - \ln(1 - 0.9)}{0.6} = 5.504, \\
\text{VaR}_{X_{(1)}}(0.9)&= \frac{-\ln(1 - 0.9)}{0.5+0.6} = 2.09~{}~{} \text{and} ~{}~{} \text{CTE}_{X_{(2)}}(0.9)= \frac{1 -\ln(1 - 0.9)}{0.5+0.6} = 3.
\end{align*}
For $ \text{VaR}_{X_{(2)}}(0.9)= Q_{0.9}$,~{} 1 - $e^{-0.5 Q_{\alpha}}- e^{-0.6 Q_{\alpha}} + e^{-1.1 Q_{\alpha}}= 0.9$. Thus, $ \text{VaR}_{X_{(2)}}(0.9)= Q_{0.9} = 5.47$. For $ \text{CTE}_{X_{(2)}}(0.9)$, we have from above expression, $ \text{CTE}_{X_{(2)}}(0.9)= 7.37. $ \\

\textbf{Case (ii)} When $X_1$ and $X_2$ are dependent, we use FGM copula $C(u,v) = uv + \theta uv(1- u)(1 -v)$ where $0 \leq u,v \leq 1$ , $-1 \leq \theta \leq 1$.
In the next section, we use this copula to analyze the dependency between the sub-portfolios.

\subsubsection{VaR and CTE for minimum of two risks of exponential distribution}
Consider $X_{(1)} = \min(X_1, X_2), u = F_{X_1}(x)$ and  $v = F_{X_2}(x)$, the tail probability distribution of $X_{(1)} = \min(X_1, X_2)$ is given by
\begin{align*}
P(X_{(1)} > x) &= P(\min (X_1, X_2) > x) = P(X_1 > x, X_2 > x),\\
&= \bar{C}(1-u, 1-v)= 1 - u - v + C(u,v), \\
& = 1 - (1 - e^{-\lambda_1 x}) - (1 - e^{-\lambda_2 x})+(1 - e^{-\lambda_1 x})(1 - e^{-\lambda_2 x})\\
&\quad + \theta (1 - e^{-\lambda_1 x})(1 - e^{-\lambda_2 x})e^{-\lambda_1 x}e^{-\lambda_2 x},\\
&= e^{-(\lambda_1+ \lambda_2)x}+ \theta e^{-(\lambda_1+ \lambda_2)x}(1 - e^{-\lambda_1 x}- e^{-\lambda_2 x}+ e^{-(\lambda_1+ \lambda_2)x}),
\end{align*}
where FGM copula $C(u,v) = uv + \theta uv(1- u)(1 -v)$ is used.\\
 So, the distribution function of  $X_{(1)}$ is given by
$$ F_{X_{(1)}}(x) = 1 - P(X_{(1)} > x) = 1 - e^{-(\lambda_1+ \lambda_2)x}-  \theta e^{-(\lambda_1+ \lambda_2)x}\big[1 - e^{-\lambda_1 x}- e^{-\lambda_2 x}+ e^{-(\lambda_1+ \lambda_2)x}\big] $$
and the probability density function of  $X_{(1)}$ is given by
\begin{align*}
f_{X_{(1)}}(x)&= (\lambda_1+ \lambda_2)e^{-(\lambda_1+ \lambda_2)x}+ \theta\big[(\lambda_1+ \lambda_2)e^{-(\lambda_1+ \lambda_2)x}- (\lambda_1+ 2\lambda_2)e^{-(\lambda_1+ 2\lambda_2)x}\\
&\quad -(2\lambda_1+ \lambda_2)e^{-(2\lambda_1+ \lambda_2)x}+ 2(\lambda_1+ \lambda_2)e^{-2(\lambda_1+ \lambda_2)x}\big].
\end{align*}
For $\text{VaR} (Q_{\alpha})$, we have
$$ F_{X_{(1)}}(x)= 1 - e^{-(\lambda_1+ \lambda_2)Q_{\alpha}}-  \theta e^{-(\lambda_1+ \lambda_2)Q_{\alpha}}\big[1 - e^{-\lambda_1 Q_{\alpha}}- e^{-\lambda_2 Q_{\alpha}}+ e^{-(\lambda_1+ \lambda_2)Q_{\alpha}}\big]= \alpha. $$
If we choose $\lambda_1 = 0.5, \lambda_2 = 0.6 $ and $\alpha = 0.9 $, then \\
$$ 1 - e^{-1.1 Q_{\alpha}}- \theta e^{-1.1 Q_{\alpha}}[1 - e^{-0.5 Q_{\alpha}}- e^{-0.6 Q_{\alpha}} + e^{-1.1 Q_{\alpha}}]= 0.9. $$
Let us take different values of $\theta$ (from weak dependency to strong dependency), we have

\begin{table}[hbt!]
\begin{center}
\maketitle
\begin{tabular}{|c|c|c|c|c|c|}
\hline
$\theta$ & 0.1 & 0.3 & 0.5 & 0.7 & 0.9\\
\hline
$\text{VaR}_{X_{(1)}}(0.9)$ & 2.14 & 2.22 & 2.3 & 2.38 & 2.45\\
\hline
\end{tabular}
\caption{\label{tab:1} Table of $\text{VaR}_{X_{(1)}}(0.9)$ vs dependency}
\end{center}
\end{table}

For $\text{CTE}_{X_{(1)}}(\alpha)$,
\begin{align*}
\text{CTE}_{X_{(1)}}(\alpha)&= E[X_{(1)}|X_{(1)} > Q_{\alpha}] = \frac{1}{1 - \alpha}\int_{Q_{\alpha}}^{\infty} x f_{X_{(1)}}(x)dx ,\\
&=  \frac{1}{1 - \alpha}\Big[\Big(Q_{\alpha}e^{-Q_{\alpha}(\lambda_1 + \lambda_2)}+ \frac{e^{-Q_{\alpha}(\lambda_1 + \lambda_2)}}{\lambda_1 + \lambda_2}\Big)+ \theta\Big(Q_{\alpha}e^{-Q_{\alpha}(\lambda_1 + \lambda_2)}+ \frac{e^{-Q_{\alpha}(\lambda_1 + \lambda_2)}}{\lambda_1 + \lambda_2}\Big)\\
&\quad - \theta\Big(Q_{\alpha}e^{-Q_{\alpha}(\lambda_1 + 2\lambda_2)}+ \frac{e^{-Q_{\alpha}(\lambda_1 + 2 \lambda_2)}}{\lambda_1 + 2\lambda_2}\Big) - \theta\Big(Q_{\alpha}e^{-Q_{\alpha}(2\lambda_1 + \lambda_2)} + \frac{e^{-Q_{\alpha}(2\lambda_1 + \lambda_2)}}{2\lambda_1 + \lambda_2}\Big)\\
&\quad+ \theta\Big(Q_{\alpha}e^{-2Q_{\alpha}(\lambda_1 + \lambda_2)}+ \frac{e^{-2Q_{\alpha}(\lambda_1 + \lambda_2)}}{2(\lambda_1 + \lambda_2)}\Big)\Big].
\end{align*}

For $\lambda_1 = 0.5$, $\lambda_1 = 0.6,$ and $\alpha = 0.9,$ we have,
\begin{align*}
\text{CTE}_{X_{(1)}}(\alpha)&= 10\Big[\Big(Q_{\alpha}e^{-1.1 Q_{\alpha}}+ \frac{Q_{\alpha}e^{-1.1 Q_{\alpha}}}{1.1}\Big) + \theta\Big(Q_{\alpha}e^{-1.1 Q_{\alpha}}+ \frac{Q_{\alpha}e^{-1.1 Q_{\alpha}}}{1.1}\Big)\\
&\quad - \theta\Big(Q_{\alpha}e^{-1.7 Q_{\alpha}} + \frac{Q_{\alpha}e^{-1.7 Q_{\alpha}}}{1.7}\Big)-  \theta\Big(Q_{\alpha}e^{-1.6 Q_{\alpha}}+ \frac{Q_{\alpha}e^{-1.6 Q_{\alpha}}}{1.6}\Big) \\
&\quad + \theta\Big(Q_{\alpha}e^{-2.2 Q_{\alpha}}+ \frac{Q_{\alpha}e^{-2.2 Q_{\alpha}}}{2.2}\Big)\Big].
\end{align*}
For different values of $\theta$, the values of $\text{CTE}_{X_{(1)}}(0.9)$ are shown in table 2.

\begin{table}[hbt!]
\begin{center}
\maketitle
\begin{tabular}{|c|c|c|c|c|c|}
\hline
$\theta$ & 0.1 & 0.3 & 0.5 & 0.7 & 0.9\\
\hline
$\text{CTE}_{X_{(1)}}(0.9)$ & 3.04 & 3.17 & 3.23 & 3.35 & 3.44\\
\hline
\end{tabular}\\
\caption{\label{tab:2} Table of  $\text{CTE}_{X_{(1)}}(0.9)$ vs dependency}
\end{center}
\end{table}

We have shown by this example that as the strength of dependency increases, VaR and CTE of the minimum of two risks also increase. Hence, it is better to make the risks $X_1$ and $X_2$ in the sub-portfolios independent or less dependent so that VaR and CTE of the minimum risks can be smaller than that for the dependent case.\\
Line graphs of $\text{VaR}_{X_{(1)}}(0.9)$ and $\text{CTE}_{X_{(1)}}(0.9)$ at different dependencies are shown in figure (1).
\begin{figure}[hbt!]
\centering
\caption{Line graphs of VaR and CTE vs dependency}
\label{Graph 5.1}
\includegraphics[scale=.7]{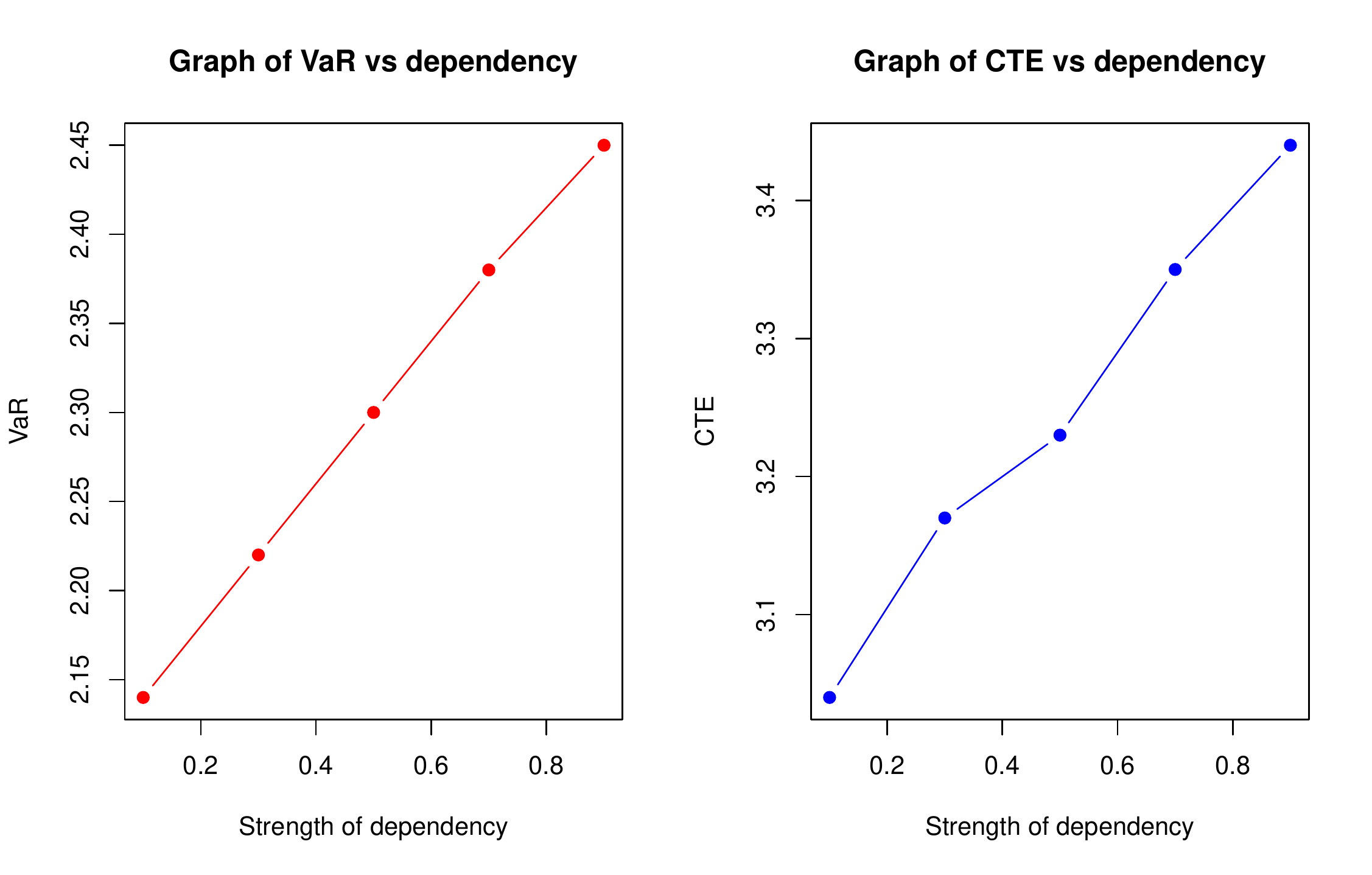}
\end{figure}
\subsubsection{VaR and CTE for maximum of two risks}
Consider $X_{(2)}= \max(X_1, X_2)$ where $X_1 \sim Exp(\lambda_1)$,  $ X_2 \sim Exp(\lambda_2)$, the distribution function of  $X_{(2)}$ is given by
$$ F_{X_{(2)}}(x)=P(X_{(2)} \leq x) = P(\max(X_1, X_2) \leq x)=  P(X_1 \leq x , X_2 \leq x)= C(u,v)$$
where $u = F_{X_1}(x) = 1 - e^{-\lambda_1 x}$ and $v = F_{X_2}(x) = 1 - e^{-\lambda_2 x}.$ Thus,
\begin{align*}
F_{X_{(2)}}(x)&= uv + \theta uv(1 - u)(1 - v),~{}~{} -1 \leq \theta \leq 1.\\
&= (1 - e^{-\lambda_1 x})(1 - e^{-\lambda_2 x})+ \theta (1 - e^{-\lambda_1 x})(1 - e^{-\lambda_2 x}).e^{-\lambda_1 x}.e^{-\lambda_2 x} \\
&= 1 - e^{-\lambda_1 x}- e^{-\lambda_2 x} + e^{-(\lambda_1 + \lambda_2)x}+ \theta\big[e^{-(\lambda_1 + \lambda_2)x}- e^{-(2\lambda_1 + \lambda_2)x}- e^{-(\lambda_1 + 2\lambda_2)x} \\
&\quad + e^{-2(\lambda_1 + \lambda_2)x}\big].
\end{align*}
Differentiating it with respect to $x$, we get the density function of $X_{(2)}$,
\begin{align*}
f_{X_{(2)}}(x) &= \lambda_1 e^{-\lambda_1 x}+ \lambda_2 e^{-\lambda_2 x} - (\lambda_1 + \lambda_2)e^{-(\lambda_1 + \lambda_2)x}+ \theta[(\lambda_1 + 2\lambda_2)e^{-(\lambda_1 + 2\lambda_2)x}+\\
&\quad (2\lambda_1 + \lambda_2)e^{-(2\lambda_1 + \lambda_2)x}- (\lambda_1 + \lambda_2)e^{-(\lambda_1 + \lambda_2)x}- 2(\lambda_1 + \lambda_2)e^{-2(\lambda_1 + \lambda_2)x}].
\end{align*}
For VaR $_{X_{(2)}}(\alpha)= Q_{\alpha}$, if we choose $\lambda_1= 0.5$, $\lambda_2 = 0.6$ and $\alpha = 0.9$, then
$$ 1 - e^{-0.5 Q_{\alpha}}- e^{-0.6 Q_{\alpha}} + e^{-1.1 Q_{\alpha}}+ \theta\big[e^{-1.1 Q_{\alpha}}- e^{-1.6 Q_{\alpha}}- e^{-1.7Q_{\alpha}}+ e^{-2.2 Q_{\alpha}}\big]= 0.9. $$
Taking different values of $\theta$ from $0.1$ to $0.9$ (weak dependency to strong dependency), we get different values of  VaR $_{X_{(2)}}(0.9)= Q_{0.9},$ which are  shown in table 3.

\begin{table}[hbt!]
\begin{center}
\maketitle
\begin{tabular}{|c|c|c|c|c|c|}
\hline
$\theta$ & 0.1 & 0.3 & 0.5 & 0.7 & 0.9\\
\hline
$VaR_{X_{(2)}}(0.9)$ & 5.46 & 5.45 & 5.45 & 5.44 & 5.43\\
\hline
\end{tabular}
\caption{\label{tab:3} Table of  VaR $_{X_{(2)}}(0.9)$ vs dependency}
\end{center}
\end{table}

For $CTE_{X_{(2)}}(0.9)$,
$$ CTE_{X_{(2)}}(\alpha)= E\big[X_{(2)}|X_{(2)} > Q_{\alpha}\big] = \frac{1}{1 - \alpha}\int_{Q_{\alpha}}^{\infty} x f_{X_{(2)}}(x)dx $$
where $ f_{X_{(2)}}(x)$ is the pdf of $ X_{(2)}= \max(X_1, X_2)$.
\begin{align*}
CTE_{X_{(2)}}(0.9)&= \frac{1}{1 - 0.9}\int_{Q_{\alpha}}^{\infty}\Big[0.5x e^{-0.5 x} + 0.6 x e^{-0.6 x}- 1.1 x e^{-1.1 x}+ \\
&\quad \theta\big(1.7 x e^{-1.7 x}+ 1.6 x e^{-1.6 x}- 1.1 x e^{-1.1 x}- 2.2 x e^{-2.2 x}\big)\Big]dx.
\end{align*}
After calculations, we have
\begin{align*}
CTE_{X_{(2)}}(0.9)&= 10\Big[\Big(Q_{\alpha} e^{-0.5 Q_{\alpha}}+ \frac{e^{-0.5 Q_{\alpha}}}{0.5}\Big)+ \Big(Q_{\alpha} e^{-0.6 Q_{\alpha}}+ \frac{e^{-0.6 Q_{\alpha}}}{0.6}\Big)- \Big(Q_{\alpha} e^{-1.1 Q_{\alpha}}\\
&\quad + \frac{e^{-1.1 Q_{\alpha}}}{1.1}\Big) + \theta\Big(\Big(Q_{\alpha} e^{-1.7 Q_{\alpha}}+ \frac{e^{-1.7 Q_{\alpha}}}{1.7}\Big)+ \Big(Q_{\alpha} e^{-1.6 Q_{\alpha}}\\
&\quad + \frac{e^{-1.6 Q_{\alpha}}}{1.6}\Big)- \Big(Q_{\alpha} e^{-1.1 Q_{\alpha}}+ \frac{e^{-1.1 Q_{\alpha}}}{1.1}\Big)- \Big(Q_{\alpha} e^{-2.2 Q_{\alpha}}+ \frac{e^{-2.2 Q_{\alpha}}}{2.2}\Big)\Big)\Big].
\end{align*}

For different values of $\theta$,  the values of $ CTE_{X_{(2)}}(0.9)$ are shown in table 4.

\begin{table}[hbt!]
\begin{center}
\maketitle
\begin{tabular}{|c|c|c|c|c|c|}
\hline
$\theta$ & 0.1 & 0.3 & 0.5 & 0.7 & 0.9\\
\hline
$CTE_{X_{(2)}}(0.9)$ & 7.369 & 7.366 & 7.361 & 7.356 & 7.351\\
\hline
\end{tabular}
\caption{\label{tab:4} Table of $ CTE_{X_{(2)}}(0.9)$ vs dependency}
\end{center}
\end{table}
We have shown by this example that as the strength of dependency increases, VaR and CTE of the maximum of two risks $ X_{(2)}= \max(X_1, X_2)$ do not change significantly. That means that, if we consider the maximum of our two investments in two different portfolios, it does not matter whether the portfolios $X_1$ and $X_2$ are dependent or not. \\

\subsection{VaR and CTE for extreme risks of Pareto distribution}
Let $X_1$ and $X_2$ follow the pareto distributions given respectively by
$ P(X_i \leq x) = F_{X_i}(x)=  1 - \Big(\dfrac{x_o}{x}\Big)^{\gamma_i}$, where $ i = 1, 2 $ and $ x_o \geq 0,~{} \gamma >0. $ \\
The extreme risks in a portfolio are $X_{(1)} = \min(X_1, X_2)$ and $X_{(2)}= \max(X_1, X_2).$\\
Then, the distribution function of $X_{(1)}= \min(X_1, X_2)$ is given by
$$ P(X_{(1)} \leq x) = 1 - P(\min(X_1, X_2) > x) = 1 - P(X_1 > x , X_2 > x). $$

\textbf{Case (i)} When $X_1$ and $X_2$ are independent, for $ X_{(1)}= \min(X_1, X_2)$, we have
$$ P(X_{(1)} \leq x) = 1 - P(X_1 > x , X_2 > x)= 1 - \left(\frac{x_o}{x}\right)^{\gamma_1}\left(\frac{x_o}{x}\right)^{\gamma_2} = 1 - \left(\frac{x_o}{x}\right)^{\gamma_1+ \gamma_2}. $$
So, $ X_{(1)} \sim \text{Pareto}(\gamma_1 + \gamma_2) $. Then, $\text{VaR}_{X_{(1)}}(\alpha)$ and  $\text{CTE}_{X_{(1)}}(\alpha)$ are given by
\begin{align*}
\text{VaR}_{X_{(1)}}(\alpha)&= x_o(1 - \alpha)^{-\frac{1}{\gamma_1+ \gamma_2}}, ~{}~{}
\text{CTE}_{X_{(1)}}(\alpha)= \frac{(\gamma_1 + \gamma_2)x_o}{(1 - \alpha)^{\frac{1}{\gamma_1 + \gamma_2}}(\gamma_1 + \gamma_2 - 1)}.
\end{align*}

For $ X_{(2)}= \max(X_1, X_2)$, the distribution function of $ X_{(2)}$  is given by
\begin{align*}
P(X_{(2)} \leq x) &= P\big(\max(X_1, X_2) \leq x\big)= P(X_1 \leq x, X_2 \leq x) \\
& = \Big[1 - \Big(\frac{x_o}{x}\Big)^{\gamma_1}\Big]\Big[1 - \Big(\frac{x_o}{x}\Big)^{\gamma_2}\Big]= 1 - \Big(\frac{x_o}{x}\Big)^{\gamma_1}- \Big(\frac{x_o}{x}\Big)^{\gamma_2}+ \Big(\frac{x_o}{x}\Big)^{\gamma_1 + \gamma_2},
\end{align*}
and its density function is given by
$$f_{X_{(2)}} =\dfrac{\gamma_1 x_o^{\gamma_1}}{x^{\gamma_1 + 1}}+ \dfrac{\gamma_2 x_o^{\gamma_2}}{x^{\gamma_2 + 1}}- \dfrac{(\gamma_1+ \gamma_2) x_o^{(\gamma_1+ \gamma_2)}}{x^{\gamma_1+ \gamma_2 + 1}}. $$

If $\text{VaR}_{X_{(2)}}(x) = Q_{\alpha}$, then $ 1 - \left(\dfrac{x_o}{x}\right)^{\gamma_1}- \left(\dfrac{x_o}{x}\right)^{\gamma_2}+ \left(\dfrac{x_o}{x}\right)^{\gamma_1 + \gamma_2}= \alpha ,$ and
\begin{align*}
\text{CTE}_{X_{(2)}}(\alpha)&= \frac{1}{1-\alpha} \int_{Q_{\alpha}}^{\infty} x f_{X_{(2)}}(x)dx, \\
& = \frac{1}{1-\alpha} \int_{Q_{\alpha}}^{\infty}\left[\frac{\gamma_1 x_o^{\gamma_1}}{x^{\gamma_1}}+ \frac{\gamma_2 x_o^{\gamma_2}}{x^{\gamma_2}}- \frac{(\gamma_1+ \gamma_2) x_o^{(\gamma_1+ \gamma_2)}}{x^{\gamma_1+ \gamma_2}}\right]dx,\\
&= \frac{1}{1-\alpha}\bigg[\frac{\gamma_1 x_o^{\gamma_1}}{(\gamma_1 - 1)(Q_{\alpha})^{\gamma_1-1}}+ \frac{\gamma_2 x_o^{\gamma_2}}{(\gamma_2 - 1)(Q_{\alpha})^{\gamma_2-1}} + \\
&\quad \frac{(\gamma_1+\gamma_2) x_o^{(\gamma_1+ \gamma_2)}}{(\gamma_1+ \gamma_2 - 1)(Q_{\alpha})^{\gamma_1+ \gamma_2 -1}}\bigg].
\end{align*}

\textbf{Example:} Assume that $x_o = 1$, $\gamma_1 = 3$, $\gamma_2 = 4$ and $\alpha = 0.9,$ then
\begin{align*}
\text{VaR}_{X_1}(0.9)&= x_o(1 - \alpha)^{-\frac{1}{\gamma_1}}= (1 - 0.9)^{-\frac{1}{3}} = 2.154, \\
\text{VaR}_{X_2}(0.9)&= x_o(1 - \alpha)^{-\frac{1}{\gamma_2}}= (1 - 0.9)^{-\frac{1}{4}} = 1.778,\\
\text{CTE}_{X_1}(0.9)&=  \frac{\gamma_1}{\gamma_1 - 1} VaR_{X_1}(0.9)=  \frac{3}{3 - 1}\times 2.154 = 3.23,\\
\text{CTE}_{X_2}(0.9)&=  \frac{\gamma_2}{\gamma_2 - 1} VaR_{X_2}(0.9)=  \frac{4}{4 - 1}\times 1.778 = 2.37 ,\\
\text{VaR}_{X_{(1)}}(0.9)&= x_o(1 - \alpha)^{-\frac{1}{(\gamma_1+ \gamma_2)}}= (1 - 0.9)^{-\frac{1}{3+4}} = 1.389,
\end{align*}
and $\text{VaR}_{X_{(2)}}(0.9)$ satisfies the expression
$$ 1 - \left(\frac{1}{Q_{\alpha}}\right)^{3}- \left(\frac{1}{Q_{\alpha}}\right)^{4}+ \left(\frac{1}{Q_{\alpha}}\right)^{7}= 0.9. $$
This gives,  $Q_{\alpha}= \text{VaR}_{X_{(2)}}(0.9)= 2.4022. $ \\
Next,
$$ \text{CTE}_{X_{(1)}}(\alpha)= \frac{(\gamma_1 + \gamma_2)x_o}{(1 - \alpha)^{\frac{1}{\gamma_1 + \gamma_2}}(\gamma_1 + \gamma_2 - 1)}= 1.62,$$
and
\begin{align*}
\text{CTE}_{X_{(2)}}(\alpha)&= \frac{1}{\alpha-1}\bigg[\frac{\gamma_1 x_o^{\gamma_1}}{(\gamma_1 - 1)(Q_{\alpha})^{\gamma_1-1}}+ \frac{\gamma_2 x_o^{\gamma_2}}{(\gamma_2 - 1)(Q_{\alpha})^{\gamma_2-1}} -
\frac{(\gamma_1+\gamma_2) x_o^{(\gamma_1+ \gamma_2)}}{(\gamma_1+ \gamma_2 - 1)(Q_{\alpha})^{\gamma_1+ \gamma_2 -1}}\bigg] = 3.5005.
\end{align*}

\textbf{Case (ii)} When $X_1$ and $X_2$ are dependent, we use FGM copula given by\\
$C(u,v) = uv + \theta uv(1- u)(1 - v),$ where  $0 \leq u,v \leq 1$, and  $-1 \leq \theta \leq 1$.

\subsubsection{VaR and CTE for minimum of two risks of Pareto distribution}
For $X_{(1)} = \min(X_1, X_2)$, $u = F_{X_1}(x)$ and $v = F_{X_2}(x)$, the survival function of $X_{(1)}$ is given by,
\begin{align*}
P(X_{(1)} > x) &= P(\min(X_1, X_2) > x)=  P(X_1 > x , X_2 > x), \\
& = \bar{C}( 1-u, 1-v)= 1 - u - v + C(u,v),\\
& = 1 - u - v + uv + \theta uv(1- u)(1 - v),\\
& = 1 - \left[1 - \left(\frac{x_o}{x}\right)^{\gamma_1}\right]- \left[1 - \left(\frac{x_o}{x}\right)^{\gamma_2}\right]+ \left[1 - \left(\frac{x_o}{x}\right)^{\gamma_1}\right] \left[1 - \left(\frac{x_o}{x}\right)^{\gamma_2}\right],\\
&\quad + \theta\left[1 - \left(\frac{x_o}{x}\right)^{\gamma_1}\right]\left[1 - \left(\frac{x_o}{x}\right)^{\gamma_2}\right]\left(\frac{x_o}{x}\right)^{\gamma_1}\left(\frac{x_o}{x}\right)^{\gamma_2}, \\
&= \left(\frac{x_o}{x}\right)^{\gamma_1+ \gamma_2}+ \theta\left[\left(\frac{x_o}{x}\right)^{\gamma_1+ \gamma_2}- \left(\frac{x_o}{x}\right)^{2\gamma_1+ \gamma_2}- \left(\frac{x_o}{x}\right)^{\gamma_1+ 2\gamma_2}+ \left(\frac{x_o}{x}\right)^{2(\gamma_1+ \gamma_2)}\right],
\end{align*}
where FGM copula $C(u,v) = uv + \theta uv(1- u)(1 -v)$ is used.\\
So, the distribution function of  $X_{(1)}$ is given by
$$ F_{X_{(1)}}(x) = 1 - \Big(\frac{x_o}{x}\Big)^{\gamma_1+ \gamma_2}- \theta\bigg[\Big(\frac{x_o}{x}\Big)^{\gamma_1+ \gamma_2}- \Big(\frac{x_o}{x}\Big)^{2\gamma_1+ \gamma_2}- \Big(\frac{x_o}{x}\Big)^{\gamma_1+ 2\gamma_2}+ \Big(\frac{x_o}{x}\Big)^{2(\gamma_1+ \gamma_2)}\bigg],$$
and the probability density function of  $X_{(1)}$ is given by
\begin{align*}
f_{X_{(1)}}(x) &= \frac{(\gamma_1+ \gamma_2)(x_o)^{\gamma_1+\gamma_2}}{x^{\gamma_1 + \gamma_2+1}} + \theta \bigg[\frac{(\gamma_1+ \gamma_2)(x_o)^{\gamma_1+\gamma_2}}{x^{\gamma_1 + \gamma_2+1}} - \frac{(2\gamma_1+ \gamma_2)(x_o)^{2\gamma_1+\gamma_2}}{x^{2\gamma_1 + \gamma_2+1}}\\
&\quad -\frac{(\gamma_1+ 2\gamma_2)(x_o)^{\gamma_1+2\gamma_2}}{x^{\gamma_1 + 2\gamma_2+1}}+ \frac{2(\gamma_1+ \gamma_2)(x_o)^{2\gamma_1+2\gamma_2}}{x^{2\gamma_1 + 2\gamma_2+1}}\bigg].
\end{align*}
To find the value of $\text{VaR}_{X_{(1)}}(\alpha)=Q_{\alpha},$ we have
\begin{align*}
F_{X_{(1)}}(x) &= 1 - \left(\frac{x_o}{Q_{\alpha}}\right)^{\gamma_1 + \gamma_2}- \theta\bigg[\left(\frac{x_o}{Q_{\alpha}}\right)^{\gamma_1 + \gamma_2}- \left(\frac{x_o}{Q_{\alpha}}\right)^{2\gamma_1 + \gamma_2}- \left(\frac{x_o}{Q_{\alpha}}\right)^{\gamma_1 + 2\gamma_2}+ \left(\frac{x_o}{Q_{\alpha}}\right)^{2(\gamma_1 + \gamma_2)}\bigg]= \alpha.
\end{align*}

If we choose $x_o = 1$, $\gamma_1= 3$, $\gamma_2= 4 $ and $ \alpha = 0.9,$ then
$$ 1 - (Q_{\alpha})^{-7} - \theta\big[(Q_{\alpha})^{-7}- (Q_{\alpha})^{-10}- (Q_{\alpha})^{-11}+ (Q_{\alpha})^{-14}\big] = 0.9. $$
For the strength of dependency, we consider the dependency from weak to strong which is shown in table 5.
\begin{table}[hbt!]
\begin{center}
\maketitle
\begin{tabular}{|c|c|c|c|c|c|}
\hline
$\theta$ & 0.1 & 0.3 & 0.5 & 0.7 & 0.9\\
\hline
$Q_{0.9}$ & 1.39 & 1.41 & 1.43 & 1.45 & 1.47\\
\hline
\end{tabular}\\
\caption{\label{tab:5} Table of $ Q_{0.9}$ vs dependency}
\end{center}
\end{table}
For $\text{CTE}_{X_{(1)}}(\alpha)$, we have
\begin{align*}
\text{CTE}_{X_{(1)}}(\alpha)&= E\big[X_{(1)}|X_{(1)} > Q_{\alpha}\big] = \frac{1}{1 - \alpha}\int_{Q_{\alpha}}^{\infty} x f_{X_{(1)}}(x)dx, \\
&= \frac{1}{1 - \alpha}\int_{Q_{\alpha}}^{\infty}\bigg[\frac{(\gamma_1+ \gamma_2)(x_o)^{\gamma_1+\gamma_2}}{x^{\gamma_1 + \gamma_2}} + \theta\Big(\frac{(\gamma_1+ \gamma_2)(x_o)^{\gamma_1+\gamma_2}}{x^{\gamma_1 + \gamma_2}} -  \\
&\quad \frac{(2\gamma_1+ \gamma_2)(x_o)^{2\gamma_1+\gamma_2}}{x^{2\gamma_1 + \gamma_2}} -\frac{(\gamma_1+ 2\gamma_2)(x_o)^{\gamma_1+2\gamma_2}}{x^{\gamma_1 + 2\gamma_2}}+ \frac{2(\gamma_1+ \gamma_2)(x_o)^{2\gamma_1+2\gamma_2}}{x^{2\gamma_1 + 2\gamma_2}}\Big)\bigg].
\end{align*}
Using $\gamma_1 = 3$, $\gamma_2 = 4$, $x_o= 1$ and $\alpha = 0.9$, we have
$$ \text{CTE}_{X_{(1)}}(0.9)= 10\bigg[\frac{7}{6(Q_{\alpha})^6} + \theta\Big(\frac{7}{6(Q_{\alpha})^6}- \frac{10}{9(Q_{\alpha})^9}- \frac{11}{10(Q_{\alpha})^{10}}+ \frac{14}{13(Q_{\alpha})^{13}}\Big)\bigg],$$
where $ Q_{\alpha}= \text{VaR}_{X_{(1)}}(\alpha). $
The values of the conditional tail expectation at different dependencies are  shown in table 6.
\begin{table}[hbt!]
\begin{center}
\maketitle
\begin{tabular}{|c|c|c|c|c|c|}
\hline
$\theta$ & 0.1 & 0.3 & 0.5 & 0.7 & 0.9\\
\hline
$\text{CTE}_{X_{(1)}}(0.9)$ & 1.63 & 1.66 & 1.69 & 1.71 & 1.74\\
\hline
\end{tabular}
\caption{\label{tab:6} Table of $\text{CTE}_{X_{(1)}}(0.9)$ vs dependency.}
\end{center}
\end{table}
The line graphs of $\text{VaR}_{X_{(1)}}(0.9)$ and $\text{CTE}_{X_{(1)}}(0.9)$ for different positive dependencies are shown in figure 2.
\begin{figure}[hbt!]
\centering
\caption{Line graphs of VaR and CTE vs dependency}
\label{Graph 5.2}
\includegraphics[scale=0.7]{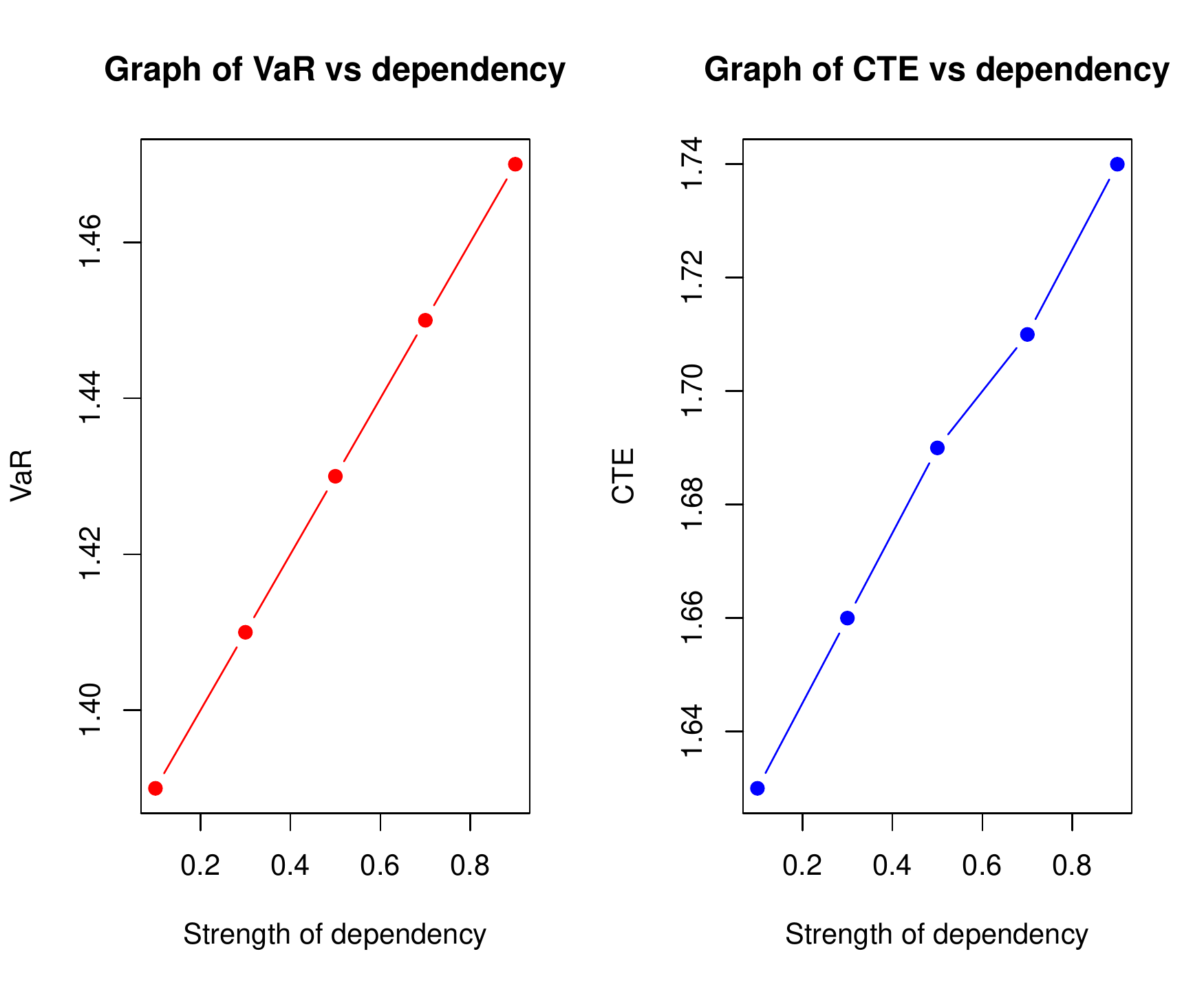}
\end{figure}

From this example, we conclude that as the strength of dependency increases, the VaR and CTE of minimum of two risks also increase for the risk of loss having the  Pareto distribution.\\
Hence, VaR and CTE of the minimum extreme risks can be smaller if we make risks $X_1$ and $X_2$ independent or less dependent.

\subsubsection{VaR and CTE for maximum of two risks of Pareto distribution}
For $X_{(2)} = \max(X_1, X_2)$, where $X_i \sim \text{Pareto}(\gamma_i)$, $i=1$,$2$, the  distribution function of $X_{(2)}$ is given by
$$ F_{X_{(2)}}(x)=  P\big(X_{(2)} \leq x\big)= P(\max(X_1, X_2) \leq x)=  P(X_1 \leq x , X_2 \leq x)= C(u,v), $$ where $C(u,v) =  uv + \theta uv(1- u)(1 - v)$,~{} $ u = F_{X_1}(x)= 1 - \Big(\dfrac{x_o}{x}\Big)^{\gamma_1}$ ~{}and ~{} $ v = F_{X_2}(x)= 1 - \Big(\dfrac{x_o}{x}\Big)^{\gamma_2}.$ \\
Then,
\begin{align*}
F_{X_{(2)}}(x)&=  uv + \theta uv(1- u)(1 - v), \\
&= \left[1 - \left(\frac{x_o}{x}\right)^{\gamma_1}\right]\left[1 - \left(\frac{x_o}{x}\right)^{\gamma_2}\right]+ \theta\left[1 - \left(\frac{x_o}{x}\right)^{\gamma_1}\right]\left[1 - \left(\frac{x_o}{x}\right)^{\gamma_2}\right]\left(\frac{x_o}{x}\right)^{\gamma_1}\left(\frac{x_o}{x}\right)^{\gamma_2}\\
&= 1 - \left(\frac{x_o}{x}\right)^{\gamma_1}- \left(\frac{x_o}{x}\right)^{\gamma_2}+ \left(\frac{x_o}{x}\right)^{\gamma_1+ \gamma_2} \\
&\quad + \theta \left[\left(\frac{x_o}{x}\right)^{\gamma_1+ \gamma_2}- \left(\frac{x_o}{x}\right)^{2\gamma_1+ \gamma_2}- \left(\frac{x_o}{x}\right)^{\gamma_1+ 2\gamma_2}+ \left(\frac{x_o}{x}\right)^{2\gamma_1+ 2\gamma_2}\right],
\end{align*}
and the probability density function of  $X_{(2)}$ is given by
\begin{align*}
f_{X_{(2)}}(x) &= \frac{\gamma_1 (x_o)^{\gamma_1}}{x^{\gamma_1 + 1}}+ \frac{\gamma_2 (x_o)^{\gamma_2}}{x^{\gamma_2 + 1}}- \frac{(\gamma_1+ \gamma_2)(x_o)^{\gamma_1+ \gamma_2}}{x^{\gamma_1 + \gamma_2 + 1}}- \theta\bigg[\frac{(\gamma_1+ \gamma_2)(x_o)^{\gamma_1+ \gamma_2}}{x^{\gamma_1 + \gamma_2 + 1}}\\
&\quad - \frac{(2\gamma_1+ \gamma_2)(x_o)^{2\gamma_1+ \gamma_2}}{x^{2\gamma_1 + \gamma_2 + 1}}- \frac{(\gamma_1+ 2\gamma_2)(x_o)^{\gamma_1+ 2\gamma_2}}{x^{\gamma_1 + 2\gamma_2 + 1}}+ \frac{(2\gamma_1+ 2\gamma_2)(x_o)^{2\gamma_1+ 2\gamma_2}}{x^{2\gamma_1 + 2\gamma_2 + 1}}\bigg].
\end{align*}

For $ \text{VaR}_{X_2}(\alpha)= Q_{\alpha} $, we choose $x_o = 1$, $\gamma_1= 3$, $\gamma_2= 4 $ and $ \alpha = 0.9.$ Then,
\begin{align*}
F_{X_{(2)}}(0.9)&= 1 - (Q_{\alpha})^{-3} - (Q_{\alpha})^{-4}+ (Q_{\alpha})^{-7} +\\
&\quad  \theta \left[(Q_{\alpha})^{-7}- (Q_{\alpha})^{-10}- (Q_{\alpha})^{-11}+ (Q_{\alpha})^{-14}\right] = 0.9.
\end{align*}
For the different strength of dependencies from weak to strong, the values of $Q_{0.9}$ are shown in  table 7.
\begin{table}[hbt!]
\begin{center}
\maketitle
\begin{tabular}{|c|c|c|c|c|c|}
\hline
$\theta$ & 0.1 & 0.3 & 0.5 & 0.7 & 0.9\\
\hline
$Q_{0.9}$ & 2.401 & 2.40 & 2.395 & 2.39 & 2.387\\
\hline
\end{tabular}
\caption{\label{tab:7} Table of $ Q_{0.9}$ vs dependency}
\end{center}
\end{table}

For $\text{CTE}_{X_{(2)}}(0.9)$, we have
$$ \text{CTE}_{X_{(2)}}(\alpha)= \frac{1}{1 - \alpha} \int_{Q_{\alpha}}^{\infty} x f_{X_{(2)}}(x)dx. $$
where $ f_{X_{(2)}}(x)$ is the pdf of $ X_{(2)}= \max(X_1, X_2)$. Hence,
\begin{align*}
\text{CTE}_{X_{(2)}}(\alpha)&=  \frac{1}{1 - \alpha} \int_{Q_{\alpha}}^{\infty} x \bigg[\frac{\gamma_1 (x_o)^{\gamma_1}}{x^{\gamma_1 + 1}}+ \frac{\gamma_2 (x_o)^{\gamma_2}}{x^{\gamma_2 + 1}}- \frac{(\gamma_1+ \gamma_2)(x_o)^{\gamma_1+ \gamma_2}}{x^{\gamma_1 + \gamma_2 + 1}} \\
&\quad - \theta[\frac{(\gamma_1+ \gamma_2)(x_o)^{\gamma_1+ \gamma_2}}{x^{\gamma_1 + \gamma_2 + 1}} - \frac{(2\gamma_1+ \gamma_2)(x_o)^{2\gamma_1+ \gamma_2}}{x^{2\gamma_1 + \gamma_2 + 1}}- \frac{(\gamma_1+ 2\gamma_2)(x_o)^{\gamma_1+ 2\gamma_2}}{x^{\gamma_1 + 2\gamma_2 + 1}}\\
&\quad + \frac{(2\gamma_1+ 2\gamma_2)(x_o)^{2\gamma_1+ 2\gamma_2}}{x^{2\gamma_1 + 2\gamma_2 + 1}}\bigg]dx.
\end{align*}

Using $x_o = 1, \gamma_1= 3, \gamma_2= 4 $ and $ \alpha = 0.9, $ and after integration, we get
\begin{align*}
\text{CTE}_{X_{(2)}}(0.9)&= 10 \bigg[\frac{3}{2(Q_{\alpha})^2}+ \frac{4}{3(Q_{\alpha})^3}- \frac{7}{6(Q_{\alpha})^6}- \theta \bigg(\frac{7}{6(Q_{\alpha})^6}- \frac{10}{9(Q_{\alpha})^9}\\
&\quad - \frac{11}{10(Q_{\alpha})^{10}}+ \frac{14}{13(Q_{\alpha})^13}\bigg)\bigg].
\end{align*}

For different strength of dependency, from weak to strong, the values of $ \text{CTE}_{X_{(2)}}(0.9)$ are shown in table 8.

\begin{table}[hbt!]
\begin{center}
\maketitle
\begin{tabular}{|c|c|c|c|c|c|}
\hline
$\theta$ & 0.1 & 0.3 & 0.5 & 0.7 & 0.9\\
\hline
$\text{CTE}_{X_{(2)}}(0.9)$ & 3.50 & 3.49 & 3.49 & 3.49 & 3.49\\
\hline
\end{tabular}
\caption{\label{tab:8} Table of $ \text{CTE}_{X_{(2)}}(0.9)$ vs dependency}
\end{center}
\end{table}

By this example, we have shown that when the strength of dependency increases, VaR and CTE of the maximum of two risks having the Pareto distribution do not change significantly. It means that VaR and CTE of the maximum of two risks do not depend on the dependency of two risks of sub-portfolios.

\subsection{VaR and CTE for aggregate risk of exponential distribution}
Let $X =(X_1, X_2)$ be a risk vector, where $X_1$ and $X_2$ denote risks or losses in sub-portfolios. Let $X = X_1+ X_2$ be the aggregate or the total risk in a portfolio. We consider, $X_i \sim \text{Exp}(\lambda_i)$, $i=1,2$ and $x>0.$ \\
Then, the distribution function of $X_i$ is given by $ F_{X_i}(x) = P(X_i \leq x_i) = 1 - e^{-\lambda_i x_i}$, $i=1, 2$, and $x > 0,$ and the distribution function of $X=X_1+X_2$ is given by
$$ P(X \leq x) = P(X_1+ X_2 \leq x).$$

\textbf{Case (i)} When $X_1$ and $X_2$ are independent, the distribution function and the density function of $X = X_1 + X_2$ are given by
$$F_{X}(x) = 1 + \frac{\lambda_1}{\lambda_2- \lambda_1}e^{-\lambda_2 x}- \frac{\lambda_2}{\lambda_2- \lambda_1}e^{-\lambda_1 x}, \text{and} f_X(x) = \frac{\lambda_1 \lambda_2}{\lambda_2- \lambda_1} (e^{-\lambda_1 x} - e^{-\lambda_2 x})$$ respectively.

\subsubsection{Value at risk (VaR) of $X=X_1+X_2$}
Let $Q_{\alpha}$ be value at risk of $X$. Then,
$$ F_X(Q_{\alpha})= 1 + \frac{\lambda_1}{\lambda_2- \lambda_1}e^{-\lambda_2 Q_{\alpha}}- \frac{\lambda_2}{\lambda_2- \lambda_1}e^{-\lambda_1 Q_{\alpha}}= \alpha $$
and the conditional tail expectation of $X$ is given as
\begin{align*}
 \text{CTE}_X(\alpha)&= \frac{1}{1- \alpha} \int_{Q_{\alpha}}^{\infty} x f_X(x)dx = \frac{1}{1- \alpha} \int_{Q_{\alpha}}^{\infty} x\left[\frac{\lambda_1 \lambda_2}{\lambda_2- \lambda_1} (e^{-\lambda_1 x} - e^{-\lambda_2 x})\right] dx \\
& = \frac{\lambda_1 \lambda_2}{(1-\alpha)(\lambda_2- \lambda_1)}\left[\frac{Q_{\alpha} e^{-\lambda_1 Q_{\alpha}}}{\lambda_1} + \frac{e^{-\lambda_1 Q_{\alpha}}}{\lambda_1^2}- \frac{Q_{\alpha} e^{-\lambda_2 Q_{\alpha}}}{\lambda_2}- \frac{e^{-\lambda_2 Q_{\alpha}}}{\lambda_2^2}\right].
\end{align*}

\textbf{Example:} If we choose $\lambda_1 = 0.5$, $\lambda_2= 0.6$ and $\alpha= 0.9$, then, value at risk $(Q_{\alpha})$ can be found from
\begin{align*}
F_X(Q_{\alpha})&= 1 + \frac{\lambda_1}{\lambda_2- \lambda_1}e^{-\lambda_2 Q_{\alpha}}- \frac{\lambda_2}{\lambda_2- \lambda_1}e^{-\lambda_1 Q_{\alpha}}= \alpha\\
&= 1+ 5e^{-0.6 Q_{\alpha}} - 6e^{-0.5 Q_{\alpha}} = 0.9.
\end{align*}
This gives, $ Q_{\alpha} = 7.14.$ Next, The conditional tail expectation of $X$ is given by
\begin{align*}
\text{CTE}_X(\alpha)&= \frac{\lambda_1 \lambda_2}{(1-\alpha)(\lambda_2- \lambda_1)}\left[\frac{Q_{\alpha} e^{-\lambda_1 Q_{\alpha}}}{\lambda_1} + \frac{e^{-\lambda_1 Q_{\alpha}}}{\lambda_1^2}- \frac{Q_{\alpha} e^{-\lambda_2 Q_{\alpha}}}{\lambda_2}- \frac{e^{-\lambda_2 Q_{\alpha}}}{\lambda_2^2}\right]\\
\text{CTE}_X(0.9)&= \frac{0.5 \times 0.6}{(1-0.9)(0.6- 0.5)}\bigg[\frac{7.14 e^{-0.5 \times 7.14}}{0.5} + \frac{e^{-0.5 \times 7.14}}{(0.5)^2}- \frac{7.14 e^{-0.6 \times 7.14}}{0.6}\\
&\quad - \frac{e^{-0.6 \times 7.14}}{(0.6)^2}\bigg]= 9.369.
\end{align*}

\textbf{Case (ii)} When the risks $X_1$ and $X_2$ are dependent, we use FGM copula $C(u,v)= \theta uv(1-u)(1-v)$, $0 \leq u,v \leq 1$, $-1 \leq \theta \leq 1$ where $u = 1- e^{-\lambda_1 x_1}$ and $v = 1 - e^{-\lambda_2 x_2}.$ \\
Then, the joint CDF of $X_1$ and $X_2$ is given by
\begin{align*}
C(u,v) &= F(x_1,x_2) = (1 - e^{-\lambda_1 x_1})(1 - e^{-\lambda_2 x_2})+ \theta(1 - e^{-\lambda_1 x_1})(1 - e^{-\lambda_2 x_2}) e^{-\lambda_1 x_1} e^{-\lambda_2 x_2} \\
&= 1 - e^{-\lambda_1 x_1}- e^{-\lambda_2 x_2}+ e^{-\lambda_1 x_1-\lambda_2 x_2}+ \theta(e^{-\lambda_1 x_1-\lambda_2 x_2}- e^{-\lambda_1 x_1-2\lambda_2 x_2}\\
&\quad - e^{-2\lambda_1 x_1-\lambda_2 x_2}+ e^{-2\lambda_1 x_1-2\lambda_2 x_2}).
\end{align*}
and its probability density function is given by
\begin{align*}
f_{X_1,X_2}(x_1,x_2) &= \lambda_1 \lambda_2  e^{-\lambda_1 x_1-\lambda_2 x_2}+ \theta\Big(\lambda_1 \lambda_2  e^{-\lambda_1 x_1-\lambda_2 x_2}- 2\lambda_1 \lambda_2  e^{-2\lambda_1 x_1-\lambda_2 x_2}\\
&\quad - 2\lambda_1 \lambda_2  e^{-\lambda_1 x_1-2\lambda_2 x_2}+ 4\lambda_1 \lambda_2  e^{-2\lambda_1 x_1-2\lambda_2 x_2}\Big).
\end{align*}
Now, the probability density function of $X= X_1+ X_2$ is given by
\begin{align*}
f_X(x) &= \int_0^x f_{X_1,X_2} (x_1, x-x_1)dx_1, \\
&= \int_0^x \Big[\lambda_1 \lambda_2  e^{-\lambda_1 x_1-\lambda_2(x-x_1)}+ \theta\Big(\lambda_1 \lambda_2 e^{-\lambda_1 x_1-\lambda_2(x-x_1)} - 2\lambda_1 \lambda_2  e^{-2\lambda_1 x_1-\lambda_2(x-x_1)}\\
&\quad- 2\lambda_1 \lambda_2  e^{-\lambda_1 x_1-2\lambda_2(x-x_1)}+ 4\lambda_1 \lambda_2 e^{-2\lambda_1 x_1-2\lambda_2(x-x_1)}\Big)\Big]dx_1.
\end{align*}
After calculations, we get
\begin{align*}
f_X(x) &= \frac{\lambda_1 \lambda_2}{\lambda_1 - \lambda_2}(e^{-\lambda_2 x} - e^{-\lambda_1 x}) + \theta\bigg[ \frac{\lambda_1 \lambda_2}{\lambda_1 - \lambda_2}(e^{-\lambda_2 x} - e^{-\lambda_1 x})- \frac{2\lambda_1 \lambda_2}{\lambda_1 - 2\lambda_2}(e^{-2\lambda_2 x}\\
&\quad - e^{-\lambda_1 x})- \frac{2\lambda_1 \lambda_2}{2\lambda_1 - \lambda_2}(e^{-\lambda_2 x} - e^{-2\lambda_1 x})+ \frac{2\lambda_1 \lambda_2}{\lambda_1 - \lambda_2}(e^{-2\lambda_2 x} - e^{-2\lambda_1 x})\bigg],
\end{align*}

and the distribution function of $X= X_1+X_2$ is given by
\begin{align*}
F_X(x) &= 1 + \frac{\lambda_1}{\lambda_2 - \lambda_1} e^{-\lambda_2 x}- \frac{\lambda_2}{\lambda_2 - \lambda_1} e^{-\lambda_1 x}+ \theta\bigg[\frac{\lambda_1}{\lambda_2 - \lambda_1} e^{-\lambda_2 x}- \frac{\lambda_2}{\lambda_2 - \lambda_1} e^{-\lambda_1 x}\\
&\quad - \frac{\lambda_1}{2\lambda_2 - \lambda_1} e^{-2\lambda_2 x} + \frac{2\lambda_2}{2\lambda_2 - \lambda_1} e^{-\lambda_1 x}- \frac{2\lambda_1}{\lambda_2 - 2\lambda_1} e^{-\lambda_2 x}\\
&\quad + \frac{\lambda_2}{\lambda_2 - 2\lambda_1}e^{-2\lambda_1 x}+ \frac{\lambda_1}{\lambda_2 - \lambda_1} e^{-2\lambda_2 x}- \frac{\lambda_2}{\lambda_2 - \lambda_1}e^{-2\lambda_1 x}\bigg].
\end{align*}
For value at risk $(VaR)$, if we choose $\lambda_1=0.5$, and $\lambda_2 = 0.6$, the distribution function of $X$ becomes
\begin{align*}
F_X(x)&= 1 + 5 e^{-0.6 x}- 6 e^{-0.5 x}+ \theta(5 e^{-0.6 x}- 6 e^{-0.5 x}- 0.714 e^{-1.2 x} + 1.714 e^{-0.5 x}\\
&\quad + 2.5 e^{-0.6 x}- 1.5 e^{-x}+ 5 e^{-1.2 x}- 6 e^{-x}).
\end{align*}
For $VaR_X(\alpha) = x = Q_{\alpha}$, we have
\begin{align*}
\alpha &= 0.9 = 1 + 5 e^{-0.6 Q_{\alpha}}- 6 e^{-0.5 Q_{\alpha}}+ \theta(5 e^{-0.6 Q_{\alpha}}- 6 e^{-0.5 Q_{\alpha}}- 0.714 e^{-1.2 Q_{\alpha}}\\
&\quad + 1.714 e^{-0.5 Q_{\alpha}}+ 2.5 e^{-0.6 Q_{\alpha}}- 1.5 e^{-Q_{\alpha}}+ 5 e^{-1.2 Q_{\alpha}}- 6 e^{-Q_{\alpha}}).
\end{align*}
We consider the different strengths of dependency from weak to strong. Then, value at risk (VaR) can be found in table 9.
\begin{table}[hbt!]
\begin{center}
\maketitle
\begin{tabular}{|c|c|c|c|c|c|}
\hline
$\theta$ & 0.1 & 0.3 & 0.5 & 0.7 & 0.9\\
\hline
$Q_{0.9}$ & 7.19 & 7.30 & 7.40 & 7.51 & 7.61\\
\hline
\end{tabular}
\caption{\label{tab:9} Table of $Q_{0.9}$ vs dependency}
\end{center}
\end{table}
From the table, we can see that as the dependency increases between the two risks in a portfolio, their aggregate value at risk also increases.

\subsubsection{Conditional tail expectation (CTE) of $X = X_1 + X_2$}

The conditional tail expectation of $X=  X_1 + X_2$ is given by
\begin{align*}
CTE_X(\alpha)&= \frac{1}{1-\alpha} \int_{Q_{\alpha}}^{\infty} x f_X(x)dx,~{} \text{where}~{}~{} Q_{\alpha}= VaR_X(\alpha).\\
&= \frac{1}{1-\alpha} \int_{Q_{\alpha}}^{\infty}\bigg[x \frac{\lambda_1 \lambda_2}{\lambda_1 - \lambda_2}(e^{-\lambda_2 x} - e^{-\lambda_1 x}) + x \theta\bigg(\frac{\lambda_1 \lambda_2}{\lambda_1 - \lambda_2}(e^{-\lambda_2 x} - e^{-\lambda_1 x}) \\
&\quad -\frac{2\lambda_1 \lambda_2}{\lambda_1 - 2\lambda_2}(e^{-2\lambda_2 x} - e^{-\lambda_1 x})- \frac{2\lambda_1 \lambda_2}{2\lambda_1 - \lambda_2}(e^{-\lambda_2 x} - e^{-2\lambda_1 x})+ \frac{2\lambda_1 \lambda_2}{\lambda_1 - \lambda_2}(e^{-2\lambda_2 x} - e^{-2\lambda_1 x})\bigg)\bigg]dx,\\
&= \frac{\lambda_1 \lambda_2}{(1-\alpha)(\lambda_1 - \lambda_2)}\bigg[\frac{Q_{\alpha}e^{-\lambda_2 Q_{\alpha}}}{\lambda_2} + \frac{e^{-\lambda_2 Q_{\alpha}}}{\lambda_2^2}- \frac{Q_{\alpha}e^{-\lambda_1 Q_{\alpha}}}{\lambda_1}-\frac{e^{-\lambda_1 Q_{\alpha}}}{\lambda_1^2} \bigg] \\
&\quad + \theta \frac{\lambda_1 \lambda_2}{(1-\alpha)(\lambda_1 - \lambda_2)}\bigg[\frac{Q_{\alpha}e^{-\lambda_2 Q_{\alpha}}}{\lambda_2} + \frac{e^{-\lambda_2 Q_{\alpha}}}{\lambda_2^2}-\frac{Q_{\alpha}e^{-\lambda_1 Q_{\alpha}}}{\lambda_1}-\frac{e^{-\lambda_1 Q_{\alpha}}}{\lambda_1^2} \bigg]\\
&\quad - \theta \frac{2\lambda_1 \lambda_2}{(1-\alpha)(\lambda_1 - 2\lambda_2)}\bigg[\frac{Q_{\alpha}e^{-2\lambda_2 Q_{\alpha}}}{2\lambda_2} + \frac{e^{-2\lambda_2 Q_{\alpha}}}{4\lambda_2^2} - \frac{Q_{\alpha}e^{-\lambda_1 Q_{\alpha}}}{\lambda_1}-\frac{e^{-\lambda_1 Q_{\alpha}}}{\lambda_1^2} \bigg]\\
&\quad - \theta \frac{2\lambda_1 \lambda_2}{(1-\alpha)(2\lambda_1 - \lambda_2)}\bigg[\frac{Q_{\alpha}e^{-\lambda_2 Q_{\alpha}}}{\lambda_2} + \frac{e^{-\lambda_2 Q_{\alpha}}}{\lambda_2^2}- \frac{Q_{\alpha}e^{-2\lambda_1 Q_{\alpha}}}{2\lambda_1} -\frac{e^{-2\lambda_1 Q_{\alpha}}}{4\lambda_1^2} \bigg]\\
&\quad + \theta \frac{2\lambda_1 \lambda_2}{(1-\alpha)(\lambda_1 - \lambda_2)}\bigg[\frac{Q_{\alpha}e^{-2\lambda_2 Q_{\alpha}}}{2\lambda_2} + \frac{e^{-2\lambda_2 Q_{\alpha}}}{4\lambda_2^2}- \frac{Q_{\alpha}e^{-2\lambda_1 Q_{\alpha}}}{2\lambda_1}-\frac{e^{-2\lambda_1 Q_{\alpha}}}{4\lambda_1^2} \bigg].
\end{align*}

Putting $\lambda_1 = 0.5, \lambda_2=0.6$ and $\alpha = 0.9$, we have
\begin{align*}
\text{CTE}_X(\alpha)&= -30\bigg[\frac{Q_{\alpha}e^{-0.6 Q_{\alpha}}}{0.6}+ \frac{e^{-0.6 Q_{\alpha}}}{0.36}- \frac{Q_{\alpha}e^{-0.5 Q_{\alpha}}}{0.5}- \frac{e^{-0.5 Q_{\alpha}}}{0.25}\bigg] -30\theta\bigg[\frac{Q_{\alpha}e^{-0.6 Q_{\alpha}}}{0.6}\\
&\quad + \frac{e^{-0.6 Q_{\alpha}}}{0.36}- \frac{Q_{\alpha}e^{-0.5 Q_{\alpha}}}{0.5}- \frac{e^{-0.5 Q_{\alpha}}}{0.25}\bigg] + 8.57\theta \bigg[\frac{Q_{\alpha}e^{-1.2 Q_{\alpha}}}{1.2}+ \frac{e^{-1.2 Q_{\alpha}}}{1.44}\\
&\quad - \frac{Q_{\alpha}e^{-0.5 Q_{\alpha}}}{0.5}- \frac{e^{-0.5 Q_{\alpha}}}{0.25}\bigg] -15\theta \bigg[\frac{Q_{\alpha}e^{-0.6 Q_{\alpha}}}{0.6}+ \frac{e^{-0.6 Q_{\alpha}}}{0.36}- Q_{\alpha}e^{-Q_{\alpha}}\\
&\quad - e^{-Q_{\alpha}}\bigg] + 60\theta \bigg[\frac{Q_{\alpha}e^{-1.2 Q_{\alpha}}}{1.2}+ \frac{e^{-1.2 Q_{\alpha}}}{1.44}- Q_{\alpha}e^{-Q_{\alpha}}- e^{-Q_{\alpha}}\bigg].
\end{align*}
Taking different strengths of dependency from weak to strong (0.1 to 0.9) and corresponding values at risk, the values of conditional tail expectation after calculations are shown in a table. Also, the line graphs of $\text{VaR}_{X}(0.9)$ and $\text{CTE}_{X}(0.9)$ for the aggregate risk of two dependent risks at different positive dependencies (from $ 0.1$ to $ 0.9$) are constructed and shown in figure 3.

\begin{table}[hbt!]
\begin{center}
\maketitle
\begin{tabular}{|c|c|c|c|c|c|}
\hline
$\theta$ & 0.1 & 0.3 & 0.5 & 0.7 & 0.9\\
\hline
$\text{CTE}_{X}(0.9)$ & 9.44 & 9.58 & 9.72 & 9.86 & 9.99\\
\hline
\end{tabular}
\caption{\label{tab:10} Table of $\text{CTE}_{X}(0.9)$ vs dependency}
\end{center}
\end{table}

\begin{figure}
\centering
\caption{Line graphs of $Q_{0.9}$ and $\text{CTE}_{X}(0.9)$ vs dependency}
\label{Graph 4.6}
\includegraphics[scale=.7]{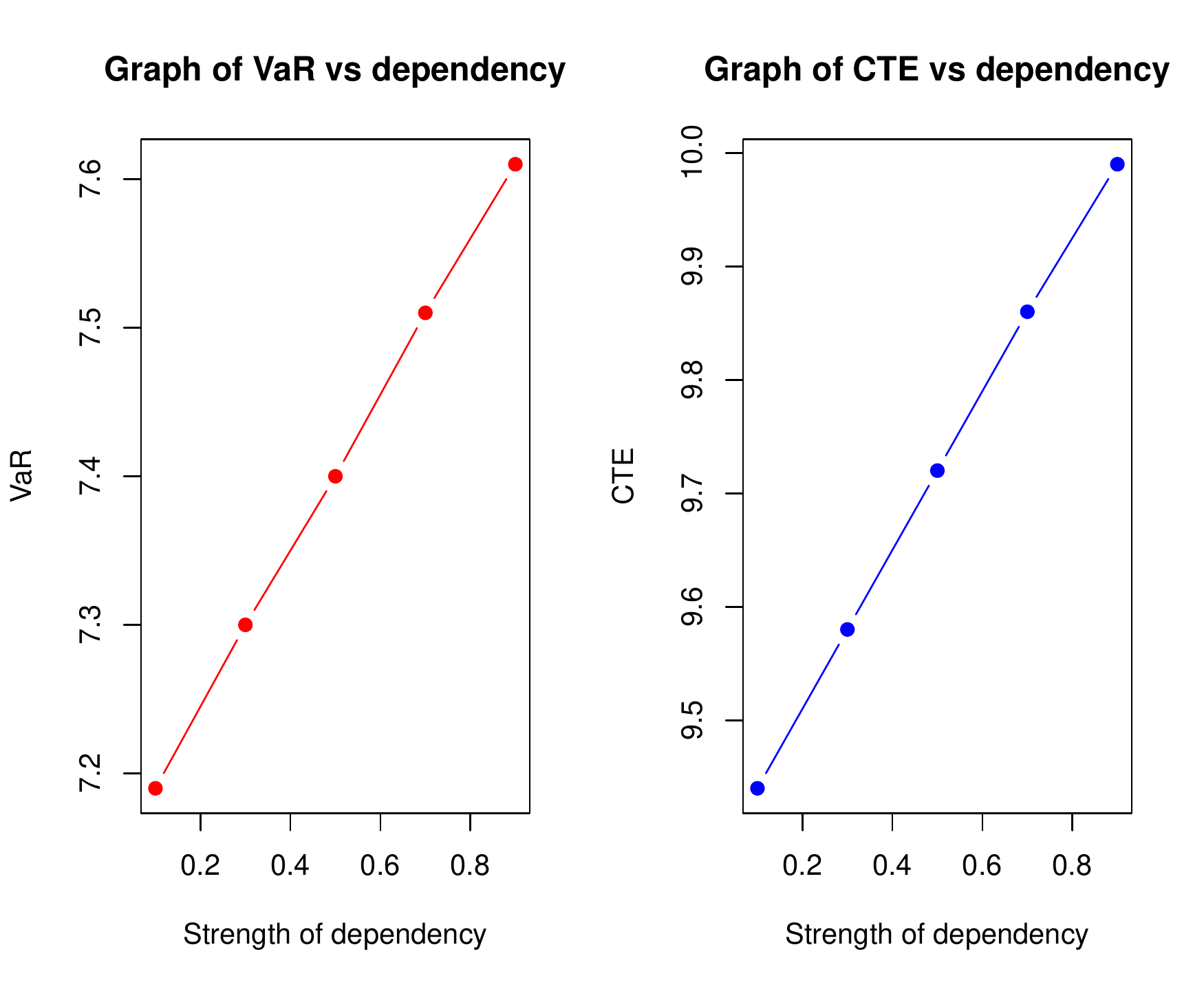}
\end{figure}

By this example, we concluded that when the strength of dependency increases, the CTE of the aggregate risk having exponential distribution increases significantly. That means that both measures, VaR and CTE of the aggregate of two risks, depend on the dependency of two risks of sub-portfolios.\\

\section{MoT  for extreme and aggregate of dependent risks}
In actuarial science, finance and insurance, several risk measures, namely, VaR, CTE, the distorted risk measures have been used in the literature. We are interested to propose an alternative risk measure which is called median of tail (MoT).
\begin{definition}
If measure of risk follows a continuous distribution $X$ with probability density function $f(x)$, distribution function $F(x)$ and having value at risk $Q_{\alpha}$, then median of tail is denoted by MoT and defined to satisfy the relation \\
$ \int_{Q_{\alpha}}^{M}f(x)dx = \dfrac{1-\alpha}{2}$, or, $F(M) = F(Q_{\alpha}) + \dfrac{1-\alpha}{2},$ with M = MoT.
\end{definition}

\subsection{MoT for exponential and pareto distribution}
Let the risk function follow the exponential distribution with pdf $f(x)= \lambda e^{-\lambda x}$,~{} $x \geq 0$, $\lambda > 0$ and CDF  $F(x) = 1- e^{-\lambda x}$, $x \geq 0$.\\
Then, the MoT (M) is given by the equation $ \int_{Q_{\alpha}}^{M} f(x)dx = \dfrac{1-\alpha}{2}$, where value at risk, ${Q_{\alpha}}= -\dfrac{\ln(1-\alpha)}{\lambda}.$ Then, the simple calculation leads to \\
$\text{MoT}_X(\alpha)= -\dfrac{1}{\lambda} {\ln \Big(\dfrac{1-\alpha}{2}\Big)}.$\\

\textbf{Remarks 1:} For $\lambda=0.5$, $\alpha=0.95$, we get, $\text{VaR}_X(0.95)= 5.99$, $\text{MoT}_X(0.95)= 7.37$, and $\text{CTE}_X(0.95)= 7.99$.

Suppose, the risk function $X$ follows the pareto distribution with pdf $f(x)= \dfrac{\gamma x_o}{x^{\gamma+1}}$, $x_o \geq 0$, $\gamma >0$ and the distribution function $F(x) = 1 - \Big(\dfrac {x_o}{x}\Big)^{\gamma}$. The MoT (M) is given by the equation $ \int_{Q_{\alpha}}^{M} f(x)dx = \dfrac{1-\alpha}{2}$.\\
Then, the simple calculation leads to
$$\text{MoT} = \text{M} = \Big[\dfrac{(1-\alpha)(x_o)^{1-\gamma}-\frac{1-\alpha}{2}}{x_o}\Big]^{-\frac{1}{\gamma}}.$$

\textbf{Remarks 2:} For $x_o=1$, $\gamma=3$,and $\alpha = 0.9$, we get, $\text{VaR}_X(0.9)= 2.15$, $\text{MoT}_X(0.9)= 2.71$, and $\text{CTE}_X(0.9)= 3.23$.

\subsection{MoT for extreme risks of exponential distribution}
Let $X_1$ and $X_2$ be $\exp(\lambda_1)$ and $\exp(\lambda_2)$ distributions. \\

\textbf{Case (i)} When $X_1$ and $X_2$ are independent, from section 3.2, the distribution function of $X_{(1)}= \min(X_1, X_2)$ is given by
$$ F_{X_{(1)}}= P(X_{(1)} \leq x) = 1 - e^{-(\lambda_1+\lambda_2)x}.$$
Thus, MoT of $X_{(1)}= \min(X_1, X_2)$ is given by
$\text{MoT}_{X_{(1)}}(\alpha)= -\dfrac{1}{\lambda_1+\lambda_2} {\ln \Big(\dfrac{1-\alpha}{2}\Big)}.$\\
Next, the distribution function of $ X_{(2)}= \max(X_1, X_2)$ is given by
$$ F_{X_{(2)}}=P(X_{(2)} \leq x) = 1 - e^{-\lambda_1 x}- e^{-\lambda_2 x}+ e^{-(\lambda_1+ \lambda)x}$$
and the density function of $ X_{(2)}= \max(X_1, X_2)$ is given by
$$ f_{X_{(2)}} = \lambda_1 e^{-\lambda_1 x} +  \lambda_2 e^{-\lambda_2 x} - (\lambda_1+ \lambda_2)  e^{-(\lambda_1+ \lambda_2)x}.$$
Then, the MoT = M of $X_{(2)} = \max(X_1, X_2),$ where $Q_{\alpha}$ is value at risk, is given by
\begin{align*}
\int_{Q_{\alpha}}^{M} f_{X_{(2)}} dx = \int_{Q_{\alpha}}^{M} \Big[\lambda_1 e^{-\lambda_1 x} +  \lambda_2 e^{-\lambda_2 x} - (\lambda_1+ \lambda_2)  e^{-(\lambda_1+ \lambda_2)x}\Big] dx = \frac{1-\alpha}{2}.
\end{align*}
After calculations, we get \\
$e^{-\lambda_1 M}+ e^{-\lambda_2 M}- e^{-(\lambda_1+ \lambda_2) M}= \dfrac{1-\alpha}{2}$, where M= MoT is median of tail.

\textbf{Examples:}
If we choose $\lambda_1 = 0.5$, $\lambda_2 = 0.6 $ and $\alpha = 0.9$, then
\begin{align*}
\text{MoT}_{X_1}(0.9)&= 5.99,~{} \text{MoT}_{X_2}(0.9) = 4.99,\\
\text{MoT}_{X_{(1)}}(0.9)&= 2.72~{} \text{and} ~{} \text{MoT}_{X_{(2)}}(0.9)= 6.78.
\end{align*}

\textbf{Case (ii)} When $X_1$ and $X_2$ are dependent, we use FGM copula $C(u,v) = uv + \theta uv(1- u)(1 -v),$ where $0 \leq u,v \leq 1$ , $-1 \leq \theta \leq 1$.
In the next subsection, we use this copula to analyze the dependency between the subportfolios.
\subsubsection{MoT for minimum of two risks of exponential distribution}
For $u = F_{X_1}(x)$ and  $v = F_{X_2}(x)$, from subsection 3.2.1, the density function of $X_{(1)} = \min(X_1, X_2)$ is given by
\begin{align*}
f_{X_{(1)}}(x)&= (\lambda_1+ \lambda_2)e^{-(\lambda_1+ \lambda_2)x}+ \theta\big[(\lambda_1+ \lambda_2)e^{-(\lambda_1+ \lambda_2)x}- (\lambda_1+ 2\lambda_2)e^{-(\lambda_1+ 2\lambda_2)x}\\
&\quad -(2\lambda_1+ \lambda_2)e^{-(2\lambda_1+ \lambda_2)x}+ 2(\lambda_1+ \lambda_2)e^{-2(\lambda_1+ \lambda_2)x}\big].
\end{align*}
Then, M= MoT of $X_{(1)} = \min(X_1, X_2)$ is given by
\begin{align*}
\dfrac{1-\alpha}{2}&=\int_{Q_{\alpha}}^{M}f_{X_{(1)}}(x)dx \\
&=\int_{Q_{\alpha}}^{M}\Big[(\lambda_1+ \lambda_2)e^{-(\lambda_1+ \lambda_2)x}+ \theta\big[(\lambda_1+ \lambda_2)e^{-(\lambda_1+ \lambda_2)x}- (\lambda_1+ 2\lambda_2)e^{-(\lambda_1+ 2\lambda_2)x}\\
&\quad -(2\lambda_1+ \lambda_2)e^{-(2\lambda_1+ \lambda_2)x}+ 2(\lambda_1+ \lambda_2)e^{-2(\lambda_1+ \lambda_2)x}\big]dx\Big].\\
&= -e^{-(\lambda_1+ \lambda_2)M}+ e^{-(\lambda_1+ \lambda_2)Q_{\alpha}}+ \theta \Big[- e^{-(\lambda_1+ \lambda_2)M}+ e^{-(\lambda_1+ \lambda_2)Q_{\alpha}}+ e^{-(\lambda_1+ 2\lambda_2)M}- e^{-(\lambda_1+ 2\lambda_2)Q_{\alpha}}\\
&\quad + e^{-(2\lambda_1+ \lambda_2)M}- e^{-(2\lambda_1+ \lambda_2)Q_{\alpha}}- e^{-2(\lambda_1+ \lambda_2)M}+ e^{-2(\lambda_1+ \lambda_2)Q_{\alpha}}\Big].
\end{align*}
For $\lambda_1=0.5$, $\lambda_2=0.6$, and $\alpha= 0.9$, we have
\begin{equation}
-e^{-1.1 M}+ e^{-1.1 Q_{\alpha}}+ \theta\Big[-e^{-1.1 M}+ e^{-1.1 Q_{\alpha}}+ e^{-1.7 M}- e^{-1.7 Q_{\alpha}}+ e^{-1.6 M}- e^{-1.6 Q_{\alpha}}- e^{-2.2 M}+ e^{-2.2 Q_{\alpha}}\Big]= 0.05.
\end{equation}
For different measures of dependency ($\theta$), and the corresponding values of value at risk  ($Q_{\alpha}$) given in table 1, the values of MoT from equation 4.1 are given in table 11.

\begin{table}[hbt!]
\begin{center}
\maketitle
\begin{tabular}{|c|c|c|c|c|c|}
\hline
$\theta$ & 0.1 & 0.3 & 0.5 & 0.7 & 0.9\\
\hline
$\text{MoT}_{X_{(1)}}(0.9)$ & 2.64 & 2.88 & 2.97 & 3.07 & 3.14 \\
\hline
\end{tabular}
\caption{\label{tab:11} Table of $\text{MoT}_{X_{(1)}}(0.9)$ vs dependency.}
\end{center}
\end{table}
From the table 11, we can see that when values of $\theta$ increase, MoT of the minimum of two risks $X_1$ and $X_2$ also increases significantly.
\subsubsection{MoT for maximum of two risks of exponential distribution}
From subsection 3.2.2, the density function of $X_{(2)} = \max(X_1, X_2)$ is given by
\begin{align*}
f_{X_{(2)}}(x) &= \lambda_1 e^{-\lambda_1 x}+ \lambda_2 e^{-\lambda_2 x} - (\lambda_1 + \lambda_2)e^{-(\lambda_1 + \lambda_2)x}+ \theta[(\lambda_1 + 2\lambda_2)e^{-(\lambda_1 + 2\lambda_2)x}+\\
&\quad (2\lambda_1 + \lambda_2)e^{-(2\lambda_1 + \lambda_2)x}- (\lambda_1 + \lambda_2)e^{-(\lambda_1 + \lambda_2)x}- 2(\lambda_1 + \lambda_2)e^{-2(\lambda_1 + \lambda_2)x}].
\end{align*}
Then, M= MoT of $X_{(2)} = \max(X_1, X_2)$ is given by
\begin{align*}
\dfrac{1-\alpha}{2}&=\int_{Q_{\alpha}}^{M}f_{X_{(2)}}(x)dx \\
&=\int_{Q_{\alpha}}^{M}\Big[\lambda_1 e^{-\lambda_1 x}+ \lambda_2 e^{-\lambda_2 x} - (\lambda_1 + \lambda_2)e^{-(\lambda_1 + \lambda_2)x}+ \theta \Big( (\lambda_1 + 2\lambda_2)e^{-(\lambda_1 + 2\lambda_2)x}+\\
&\quad (2\lambda_1 + \lambda_2)e^{-(2\lambda_1 + \lambda_2)x}- (\lambda_1 + \lambda_2)e^{-(\lambda_1 + \lambda_2)x}- 2(\lambda_1 + \lambda_2)e^{-2(\lambda_1 + \lambda_2)x}\Big)\Big]dx.\\
&= - e^{-\lambda_1 M} + e^{-\lambda_1 Q_{\alpha}} - e^{-\lambda_2 M} + e^{-\lambda_2 Q_{\alpha}}+ e^{-(\lambda_1+\lambda_2)M} - e^{-(\lambda_1+ \lambda_2)Q_{\alpha}}+ \theta \Big[-e^{-(\lambda_1+2\lambda_2)M} \\
&\quad + e^{-(\lambda_1+ 2\lambda_2)Q_{\alpha}}- e^{-(2\lambda_1+\lambda_2)M} + e^{-(2\lambda_1+ \lambda_2)Q_{\alpha}}+ e^{-(\lambda_1+\lambda_2)M} - e^{-(\lambda_1+ \lambda_2)Q_{\alpha}}+  e^{-2(\lambda_1+\lambda_2)M}\\
&\quad - e^{-2(\lambda_1+ \lambda_2)Q_{\alpha}}\Big].
\end{align*}
For $\lambda_1=0.5$, $\lambda_2=0.6$, and $\alpha= 0.9$, we have
\begin{align}
&-e^{-0.5 M}+ e^{-0.5 Q_{\alpha}}- e^{-0.6 M}+ e^{-0.6 Q_{\alpha}}+ e^{-1.1 M}- e^{-1.1 Q_{\alpha}}+ \theta\Big[-e^{-1.7 M}+ e^{-1.7 Q_{\alpha}}- e^{-1.6 M}+ e^{-1.6 Q_{\alpha}}\nonumber \\
&\quad + e^{-1.1 M}- e^{-1.1 Q_{\alpha}} + e^{-2.2 M}- e^{-2.2 Q_{\alpha}}\Big]= 0.05.
\end{align}
For different measures of dependency ($\theta$), and the corresponding values of value at risk  ($Q_{\alpha}$) given in table 2, the values of MoT from equation 4.2 are given in table 12.
\begin{table}[hbt!]
\begin{center}
\maketitle
\begin{tabular}{|c|c|c|c|c|c|}
\hline
$\theta$ & 0.1 & 0.3 & 0.5 & 0.7 & 0.9\\
\hline
$MoT_{X_{(2)}}(0.9)$ & 6.77 & 6.77 & 6.78 & 6.77 & 6.77 \\
\hline
\end{tabular}
\caption{\label{tab:12} Table of $MoT_{X_{(2)}}(0.9)$ vs dependency.}
\end{center}
\end{table}
From the table 12, we can see that when values of $\theta$ increase, MoT of the maximum of two risks $X_1$ and $X_2$ does not change significantly.

\subsection{MoT for extreme risks of Pareto distribution}
Let $X_1$ and $X_2$ follow the pareto distributions given by
$ P(X_i \leq x) = F_{X_i}(x)=  1 - \Big(\dfrac{x_o}{x}\Big)^{\gamma_i}$, where $ x_o \geq 0,~{} \gamma_i >0$, $ i = 1, 2. $ \\

\textbf{Case (i)} When $X_1$ and $X_2$ are independent, from section 3.4, the distribution function of $X_{(1)}= \min(X_1, X_2)$ is given by
$$ F_{X_{(1)}}= 1 - \left(\frac{x_o}{x}\right)^{\gamma_1}\left(\frac{x_o}{x}\right)^{\gamma_2} = 1 - \left(\frac{x_o}{x}\right)^{\gamma_1+ \gamma_2}.$$
Thus, MoT of $X_{(1)}= \min(X_1, X_2)$ is given by
$$ \text{MoT}_{X_{(1)}}(\alpha) = \Big[\dfrac{(1-\alpha)(x_o)^{1-\gamma_1-\gamma_2}-\frac{1-\alpha}{2}}{x_o}\Big]^{-\frac{1}{\gamma_1+\gamma_2}}.$$
Next, the distribution function of $ X_{(2)}= \max(X_1, X_2)$ is given by
$$ F_{X_{(2)}} = \Big[1 - \Big(\frac{x_o}{x}\Big)^{\gamma_1}\Big]\Big[1 - \Big(\frac{x_o}{x}\Big)^{\gamma_2}\Big]= 1 - \Big(\frac{x_o}{x}\Big)^{\gamma_1}- \Big(\frac{x_o}{x}\Big)^{\gamma_2}+ \Big(\frac{x_o}{x}\Big)^{\gamma_1 + \gamma_2},$$
and the density function of $ X_{(2)}= \max(X_1, X_2)$ is given by \\
$$f_{X_{(2)}} =\dfrac{\gamma_1 x_o^{\gamma_1}}{x^{\gamma_1 + 1}}+ \dfrac{\gamma_2 x_o^{\gamma_2}}{x^{\gamma_2 + 1}}- \dfrac{(\gamma_1+ \gamma_2) x_o^{(\gamma_1+ \gamma_2)}}{x^{\gamma_1+ \gamma_2 + 1}}. $$
Then, MoT = M of $X_{(2)} = \max(X_1, X_2)$ where $Q_{\alpha}$ is value at risk is given by
\begin{align*}
\int_{Q_{\alpha}}^{M} f_{X_{(2)}} dx &= \int_{Q_{\alpha}}^{M} \Big[\dfrac{\gamma_1 x_o^{\gamma_1}}{x^{\gamma_1 + 1}}+ \dfrac{\gamma_2 x_o^{\gamma_2}}{x^{\gamma_2 + 1}}- \dfrac{(\gamma_1+ \gamma_2) x_o^{(\gamma_1+ \gamma_2)}}{x^{\gamma_1+ \gamma_2 + 1}}\Big]dx = \frac{1-\alpha}{2}.\\
&=\Big(\frac{x_o}{Q_{\alpha}}\Big)^{\gamma_1} - \Big(\frac{x_o}{M}\Big)^{\gamma_1} + \Big(\frac{x_o}{Q_{\alpha}}\Big)^{\gamma_2} - \Big(\frac{x_o}{M}\Big)^{\gamma_2} - \Big(\frac{x_o}{Q_{\alpha}}\Big)^{\gamma_1+\gamma_2} + \Big(\frac{x_o}{M}\Big)^{\gamma_1+\gamma_2}.
\end{align*}
\textbf{Examples:}  For $x_o = 1$, $\gamma_1 = 3$, $\gamma_2 = 4$ and $\alpha = 0.9,$ then
\begin{align*}
\text{MoT}_{X_1}(0.9)&= 2.71,~{} \text{MoT}_{X_2}(0.9) = 2.11,\\
\text{MoT}_{X_{(1)}}(0.9)&= 1.53 ~{}~{} \text{and} ~{}~{} \text{MoT}_{X_{(2)}}(0.9)= 2.98.
\end{align*}

\textbf{Case (ii)} When $X_1$ and $X_2$ are dependent, we use FGM copula $C(u,v) = uv + \theta uv(1- u)(1 -v)$ where $0 \leq u,v \leq 1$ , $-1 \leq \theta \leq 1$.
In the next subsection, we use this copula to analyze the dependency between the subportfolios.

\subsubsection{MoT for minimum of two risks of pareto distribution}
For $u = F_{X_1}(x)$ and  $v = F_{X_2}(x)$, from subsection 3.4.1, the density function of $X_{(1)} = \min(X_1, X_2)$ is given by
\begin{align*}
f_{X_{(1)}}(x) &= \frac{(\gamma_1+ \gamma_2)(x_o)^{\gamma_1+\gamma_2}}{x^{\gamma_1 + \gamma_2+1}} + \theta \bigg[\frac{(\gamma_1+ \gamma_2)(x_o)^{\gamma_1+\gamma_2}}{x^{\gamma_1 + \gamma_2+1}} - \frac{(2\gamma_1+ \gamma_2)(x_o)^{2\gamma_1+\gamma_2}}{x^{2\gamma_1 + \gamma_2+1}}\\
&\quad -\frac{(\gamma_1+ 2\gamma_2)(x_o)^{\gamma_1+2\gamma_2}}{x^{\gamma_1 + 2\gamma_2+1}}+ \frac{2(\gamma_1+ \gamma_2)(x_o)^{2\gamma_1+2\gamma_2}}{x^{2\gamma_1 + 2\gamma_2+1}}\bigg].
\end{align*}
Then, M= MoT of $X_{(1)} = \min(X_1, X_2)$ is given by
\begin{align*}
\dfrac{1-\alpha}{2}&=\int_{Q_{\alpha}}^{M}f_{X_{(1)}}(x)dx = \dfrac{1-\alpha}{2}=\int_{Q_{\alpha}}^{M}\Big[\frac{(\gamma_1+ \gamma_2)(x_o)^{\gamma_1+\gamma_2}}{x^{\gamma_1 + \gamma_2+1}} + \theta \Big(\frac{(\gamma_1+ \gamma_2)(x_o)^{\gamma_1+\gamma_2}}{x^{\gamma_1 + \gamma_2+1}} - \frac{(2\gamma_1+ \gamma_2)(x_o)^{2\gamma_1+\gamma_2}}{x^{2\gamma_1 + \gamma_2+1}}\\
&\quad -\frac{(\gamma_1+ 2\gamma_2)(x_o)^{\gamma_1+2\gamma_2}}{x^{\gamma_1 + 2\gamma_2+1}}+ \frac{2(\gamma_1+ \gamma_2)(x_o)^{2\gamma_1+2\gamma_2}}{x^{2\gamma_1 + 2\gamma_2+1}}\Big)\Big]dx.\\
&= \Big(\dfrac{x_o}{Q_{\alpha}}\Big)^{\gamma_1+\gamma_2} - \Big(\dfrac{x_o}{M}\Big)^{\gamma_1+\gamma_2}+ \theta \Big[\Big(\dfrac{x_o}{Q_{\alpha}}\Big)^{\gamma_1+\gamma_2} - \Big(\dfrac{x_o}{M}\Big)^{\gamma_1+\gamma_2} - \Big(\dfrac{x_o}{Q_{\alpha}}\Big)^{2\gamma_1+\gamma_2} + \Big(\dfrac{x_o}{M}\Big)^{2\gamma_1+\gamma_2} \\
&\quad - \Big(\dfrac{x_o}{Q_{\alpha}}\Big)^{\gamma_1+2\gamma_2} + \Big(\dfrac{x_o}{M}\Big)^{\gamma_1+2\gamma_2} + \Big(\dfrac{x_o}{Q_{\alpha}}\Big)^{2\gamma_1+2\gamma_2} - \Big(\dfrac{x_o}{M}\Big)^{2\gamma_1+2\gamma_2}\Big].
\end{align*}
For $x_o=1$, $\gamma_1=3$, $\gamma_2=4$, $\alpha=0.9$, we have \\
\begin{align}
(Q_{\alpha})^{-7}- (M)^{-7} + \theta \Big[(Q_{\alpha})^{-7}- (M)^{-7} - (Q_{\alpha})^{-10}+ (M)^{-10}- (Q_{\alpha})^{-11}+ (M)^{-11} + (Q_{\alpha})^{-14} \nonumber \\
\quad -(M)^{-14}\Big]= 0.05.
\end{align}
For different measures of dependency ($\theta$), and the corresponding values of value at risk  ($Q_{\alpha}$) given in table 3, the values of MoT from equation 4.3 are given in table 13.

\begin{table}[hbt!]
\begin{center}
\maketitle
\begin{tabular}{|c|c|c|c|c|c|}
\hline
$\theta$ & 0.1 & 0.3 & 0.5 & 0.7 & 0.9\\
\hline
$\text{MoT}_{X_{(1)}}(0.9)$ & 1.52 & 1.55 & 1.58 & 1.61 & 1.64 \\
\hline
\end{tabular}
\caption{\label{tab:13} Table of $\text{MoT}_{X_{(1)}}(0.9)$ vs dependency.}
\end{center}
\end{table}
From table 13, we can see that when values of $\theta$ increase, MoT of the minimum of two risks $X_1$ and $X_2$ also increases significantly.

\subsubsection{MoT for maximum of two risks of pareto distribution}
From subsection 3.4.2, the density function of $X_{(2)} = \max(X_1, X_2)$ is given by
\begin{align*}
f_{X_{(2)}}(x) &= \frac{\gamma_1 (x_o)^{\gamma_1}}{x^{\gamma_1 + 1}}+ \frac{\gamma_2 (x_o)^{\gamma_2}}{x^{\gamma_2 + 1}}- \frac{(\gamma_1+ \gamma_2)(x_o)^{\gamma_1+ \gamma_2}}{x^{\gamma_1 + \gamma_2 + 1}}- \theta\bigg[\frac{(\gamma_1+ \gamma_2)(x_o)^{\gamma_1+ \gamma_2}}{x^{\gamma_1 + \gamma_2 + 1}}\\
&\quad - \frac{(2\gamma_1+ \gamma_2)(x_o)^{2\gamma_1+ \gamma_2}}{x^{2\gamma_1 + \gamma_2 + 1}}- \frac{(\gamma_1+ 2\gamma_2)(x_o)^{\gamma_1+ 2\gamma_2}}{x^{\gamma_1 + 2\gamma_2 + 1}}+ \frac{(2\gamma_1+ 2\gamma_2)(x_o)^{2\gamma_1+ 2\gamma_2}}{x^{2\gamma_1 + 2\gamma_2 + 1}}\bigg].
\end{align*}
Then, M= MoT of $X_{(2)} = \max(X_1, X_2)$ is given by
\begin{align*}
\dfrac{1-\alpha}{2}&=\int_{Q_{\alpha}}^{M}\bigg[ \frac{\gamma_1 (x_o)^{\gamma_1}}{x^{\gamma_1 + 1}}+ \frac{\gamma_2 (x_o)^{\gamma_2}}{x^{\gamma_2 + 1}}- \frac{(\gamma_1+ \gamma_2)(x_o)^{\gamma_1+ \gamma_2}}{x^{\gamma_1 + \gamma_2 + 1}}- \theta \bigg(\frac{(\gamma_1+ \gamma_2)(x_o)^{\gamma_1+ \gamma_2}}{x^{\gamma_1 + \gamma_2 + 1}}\\
&\quad - \frac{(2\gamma_1+ \gamma_2)(x_o)^{2\gamma_1+ \gamma_2}}{x^{2\gamma_1 + \gamma_2 + 1}}- \frac{(\gamma_1+ 2\gamma_2)(x_o)^{\gamma_1+ 2\gamma_2}}{x^{\gamma_1 + 2\gamma_2 + 1}}+ \frac{(2\gamma_1+ 2\gamma_2)(x_o)^{2\gamma_1+ 2\gamma_2}}{x^{2\gamma_1 + 2\gamma_2 + 1}}\bigg)\bigg]dx. \\
\end{align*}
For $x_o=1$, $\gamma_1=3$, $\gamma_2=4$, $\alpha=0.9$, we have \\
\begin{align}
(Q_{\alpha})^{-3}- (M)^{-3} + (Q_{\alpha})^{-4}- (M)^{-4} -(Q_{\alpha})^{-7} + (M)^{-7} -\theta \Big[(Q_{\alpha})^{-7}- (M)^{-7} - (Q_{\alpha})^{-10} \nonumber \\
\quad + (M)^{-10}- (Q_{\alpha})^{-11}+ (M)^{-11} + (Q_{\alpha})^{-14} -(M)^{-14}\Big]= 0.05.
\end{align}
For different strengths of dependency ($\theta$), and the corresponding values of value at risk  ($Q_{\alpha}$) given in table 3, the values of MoT from equation 4.4 are given in table 14.\\

\begin{table}[hbt!]
\begin{center}
\maketitle
\begin{tabular}{|c|c|c|c|c|c|}
\hline
$\theta$ & 0.1 & 0.3 & 0.5 & 0.7 & 0.9\\
\hline
$\text{MoT}_{X_{(2)}}(0.9)$ & 2.98 & 2.98 & 2.975 & 2.97 & 2.97 \\
\hline
\end{tabular}
\caption{\label{tab:14} Table of $\text{MoT}_{X_{(2)}}(0.9)$ vs dependency.}
\end{center}
\end{table}

From the table 14, we can see that when $\theta$ increases, MoT of the minimum of two risks $X_1$ and $X_2$  does not change significantly.

\subsection{MoT for aggregate risk of exponential distribution}
In this section, we derive MoT for the aggregate risk $X=X_1+X_2,$ where the distribution function of $X_i$ is given by 		
$ F_{X_i}(x) = P(X_i \leq x_i) = 1 - e^{-\lambda_i x_i}$, $i=1, 2$, and $x > 0.$

\textbf{Case (i)} When $X_1$ and $X_2$ are independent, from section 3.5, the distribution function and the density function of $X = X_1 + X_2$ are given by
$$F_{X}(x) = 1 + \frac{\lambda_1}{\lambda_2- \lambda_1}e^{-\lambda_2 x}- \frac{\lambda_2}{\lambda_2- \lambda_1}e^{-\lambda_1 x} ~{}~{} \text{and} ~{}~{} f_X(x) = \frac{\lambda_1 \lambda_2}{\lambda_2- \lambda_1} (e^{-\lambda_1 x} - e^{-\lambda_2 x})$$ respectively.\\
Then, M = MoT of $X = X_1 + X_2$ is given by \\
\begin{align*}
\dfrac{1-\alpha}{2}&=\int_{Q_{\alpha}}^{M}\Big[\frac{\lambda_1 \lambda_2}{\lambda_2- \lambda_1} (e^{-\lambda_1 x} - e^{-\lambda_2 x})\Big]dx = \dfrac{\lambda_1 e^{-\lambda_2 M} - \lambda_2 e^{-\lambda_1 M}- \lambda_1 e^{-\lambda_2 Q_{\alpha}} + \lambda_2 e^{-\lambda_1 Q_{\alpha}}}{\lambda_2- \lambda_1}.
\end{align*}

\textbf{Example:} If we choose $\lambda_1 = 0.5$, $\lambda_2= 0.6$ and $\alpha= 0.9$, then, MoT = 8.71.

\textbf{Case (ii)} When $X_1$ and $X_2$ are dependent, we use FGM copula $C(u,v) = uv + \theta uv(1- u)(1 -v)$ to analyze the dependency where $u = 1- e^{-\lambda_1 x_1}$ and $v = 1 - e^{-\lambda_2 x_2}.$ \\
Then, from subsection 3.5.1, the probability density function of $X=X_1+X_2$ is given by
\begin{align*}
f_X(x) &= \frac{\lambda_1 \lambda_2}{\lambda_1 - \lambda_2}(e^{-\lambda_2 x} - e^{-\lambda_1 x}) + \theta\bigg[ \frac{\lambda_1 \lambda_2}{\lambda_1 - \lambda_2}(e^{-\lambda_2 x} - e^{-\lambda_1 x})- \frac{2\lambda_1 \lambda_2}{\lambda_1 - 2\lambda_2}(e^{-2\lambda_2 x}\\
&\quad - e^{-\lambda_1 x})- \frac{2\lambda_1 \lambda_2}{2\lambda_1 - \lambda_2}(e^{-\lambda_2 x} - e^{-2\lambda_1 x})+ \frac{2\lambda_1 \lambda_2}{\lambda_1 - \lambda_2}(e^{-2\lambda_2 x} - e^{-2\lambda_1 x})\bigg].
\end{align*}
Then, M = MoT of the aggregate risk $X$ is given by \\
\begin{align*}
&= \int_{Q_{\alpha}}^{M} \bigg[ \frac{\lambda_1 \lambda_2}{\lambda_1 - \lambda_2}(e^{-\lambda_2 x} - e^{-\lambda_1 x}) + \theta\Big( \frac{\lambda_1 \lambda_2}{\lambda_1 - \lambda_2}(e^{-\lambda_2 x} - e^{-\lambda_1 x})- \frac{2\lambda_1 \lambda_2}{\lambda_1 - 2\lambda_2}(e^{-2\lambda_2 x} \\
&\quad - e^{-\lambda_1 x})- \frac{2\lambda_1 \lambda_2}{2\lambda_1 - \lambda_2}(e^{-\lambda_2 x} - e^{-2\lambda_1 x})+ \frac{2\lambda_1 \lambda_2}{\lambda_1 - \lambda_2}(e^{-2\lambda_2 x} - e^{-2\lambda_1 x})\Big)\bigg]dx. \\
&= \dfrac{\lambda_1 e^{-\lambda_2 M} - \lambda_2 e^{-\lambda_1 M}- \lambda_1 e^{-\lambda_2 Q_{\alpha}} + \lambda_2 e^{-\lambda_1 Q_{\alpha}}}{\lambda_2- \lambda_1} + \theta \bigg[ \dfrac{\lambda_1 e^{-\lambda_2 M} - \lambda_2 e^{-\lambda_1 M}- \lambda_1 e^{-\lambda_2 Q_{\alpha}} + \lambda_2 e^{-\lambda_1 Q_{\alpha}}}{\lambda_2- \lambda_1}\\
&\quad - \dfrac{\lambda_1 e^{-2\lambda_2 M} - 2\lambda_2 e^{-\lambda_1 M}- \lambda_1 e^{-2\lambda_2 Q_{\alpha}} + 2\lambda_2 e^{-\lambda_1 Q_{\alpha}}}{2\lambda_2- \lambda_1}- \dfrac{2\lambda_1 e^{-\lambda_2 M} - \lambda_2 e^{-2\lambda_1 M}- 2\lambda_1 e^{-\lambda_2 Q_{\alpha}} + \lambda_2 e^{-2\lambda_1 Q_{\alpha}}}{\lambda_2- 2\lambda_1} \\
&\quad + \dfrac{\lambda_1 e^{-2\lambda_2 M} - \lambda_2 e^{-2\lambda_1 M}- \lambda_1 e^{-2\lambda_2 Q_{\alpha}} + \lambda_2 e^{-2\lambda_1 Q_{\alpha}}}{\lambda_2- \lambda_1}\bigg]=\dfrac{1-\alpha}{2}.
\end{align*}
For $\lambda_1 = 0.5$, $\lambda_2= 0.6$ and $\alpha= 0.9$, we get
\begin{align}
& (5 + 7.5\theta)e^{-0.6 M} -(6 + 4.286 \theta)e^{-0.5 M} + 4.286 \theta e^{-1.2 M} -  7.5 \theta e^{-M} - (5 + 7.5\theta) e^{-0.6 Q_{\alpha}} \nonumber \\
&\quad +  (6 + 4.286 \theta) e^{-0.5 Q_{\alpha}} - 4.286\theta e^{-1.2 Q_{\alpha}} + 7.5 \theta e^{-Q_{\alpha}} - 0.05 = 0.
\end{align}

For different strengths of dependency ($\theta$), and the corresponding values of value at risk  ($Q_{\alpha}$) given in table 9, the values of MoT from equation (3.5) are given in table 15.\\

\begin{table}[hbt!]
\begin{center}
\maketitle
\begin{tabular}{|c|c|c|c|c|c|}
\hline
$\theta$ & 0.1 & 0.3 & 0.5 & 0.7 & 0.9\\
\hline
$\text{MoT}_{X}(0.9)$ & 8.78 & 8.93 & 9.05 & 9.20 & 9.31 \\
\hline
\end{tabular}
\caption{\label{tab:15} Table of $\text{MoT}_{X}(0.9)$ vs dependency.}
\end{center}
\end{table}

From table 15, we can see that when $\theta$ increases, MoT of the aggregate of two risks $X_1$ and $X_2$  also increases significantly.

\section{Conclusion}
Using FGM copula to capture the dependence between risk measures in the sub-portfolios, we derived the explicit expressions of value at risk and the conditional tail expectation (CTE) for the extreme risks and the aggregate risk of the portfolio. Both VaR and the CTE measure the right tail risk, which are frequently used in the insurance and financial investment. Moreover, we proposed an alternate risk measure "median of tail" (MoT). To evaluate the extreme (maximum and the minimum) of two risks, we considered the cases, where the risks follow the exponential and pareto distributions whereas for the aggregate risk, the risks follow exponential distribution.  We have shown, with examples, that as the dependency between two risks in their sub-portfolios increases:\\
(i) VaR, CTE and MoT of the minimum of the two risks and the aggregate risk are also increasing and \\
(ii) VaR, CTE and MoT of the maximum of the two risks are not changing significantly. \\
If people are interested in investing money in two or more different sub-portfolios (areas), it is better to have the areas independent (or less dependent) so that the risk measures are smaller than that for the dependent case.\\

\textbf{Acknowledgement}\\
The authors would like to acknowledge that this work is supported, in part, by the Natural Sciences and Engineering Research Council of Canada (NSERC) through a Discovery Research Grant, and Carleton University.\\

\textbf{Appendix}\\

\textbf{Introduction to copula}\\

The word copula originally came from the latin word copulare, which means to join together. In many cases of modeling, it is important to obtain the joint probability density function between two or more random variables. Even though the marginals of each of the dependent random variables are known, their joint distributions cannot, in general, be derived from their marginal distributions. The following definitions can be found in many references, for example, from the book\cite{nelsen2007introduction} by Nelsen.\\

\textbf{Mathematical definition of copula:}\\
 A two dimensional copula is a function $C :[0,1] \times [0,1] \to [0,1],$ or $C :I^2\to I$, satisfing the following two conditions:
\begin{description}
\item[(i)]  Boundary conditions:
$C(u,0)=0, C(0,v)=0,$
$C(u,1)=u, C(1,v) = v$ for all $u,v \in [0,1].$
\item[(ii)] 2-increasing property:
\vspace*{3mm}
For every $u_1$, $u_2$, $v_1$, $v_2 \in I$ such that $u_1 \leq u_2$ and $v_1 \leq v_2,$\\
$C(u_1,v_1) + C(u_2,v_2) - C(u_1,v_2) - C(u_2,v_1) \geq 0.$\\

\textbf{Note}: If $C(u,v)$ is twice differentiable, then the 2-increasing property is equivalent to $\dfrac{\partial^2 C(u,v)}{\partial u \partial v} \geq 0.$
\end{description}

\textbf{Farlie-Gumbel-Morgenstern copula:}\\
The Farlie-Gumbel-Morgenstern copula (FGM) is defined by
$$C(u,v;\theta)= uv + \theta uv(1-u)(1-v), ~{}~{} -1 \leq \theta \leq 1. $$
The FGM copula was first proposed by Morgenstern (1956). It is a perturbation of the product copula. This copula \cite{sriboonchitta2018fgm} is only useful when dependence between the two marginals is modest in magnitude.

\bibliographystyle{ieeetran}
\bibliography{refpaper2}

\end{document}